\documentstyle[aps,prc,epsf]{revtex}

\def\be{\begin{equation}}
\def\ee{\end{equation}}
\def\bea{\begin{eqnarray}}
\def\eea{\end{eqnarray}}

\begin{document}

\title{\LARGE {\bf 
B-Spline Finite Elements and their Efficiency in
Solving Relativistic Mean Field Equations
} }

\author{ W. P\"oschl \\
Physics-Department of the Duke University,\\
Durham, NC-27708, USA }
\vspace{15mm}
\date{\today}
\maketitle
\vspace{15mm}


%
\abstract{
%
A finite element method using B-splines is presented and compared with a 
conventional finite element method of Lagrangian type.
The efficiency of both methods has been investigated at the example of
a coupled non-linear system of Dirac eigenvalue equations and
inhomogeneous Klein-Gordon equations which describe a nuclear system
in the framework of relativistic mean field theory. Although, FEM
has been applied with great success in nuclear RMF recently, a well
known problem is the appearance of spurious solutions in the spectra
of the Dirac equation. The question, whether B-splines lead to a reduction
of spurious solutions is analyzed. Numerical expenses, precision and
behavior of convergence are compared for both methods in view of
their use in large scale computation on FEM grids with more dimensions. 
A B-spline version of the object oriented C++ code for spherical nuclei 
has been used for this investigation.
}
%
%
%

\centerline{\Large \bf PROGRAM SUMMARY}
\vspace{1 cm}
\begin{itemize}
\item[]{\it Title of program\/}:~ bspFEM.cc\hfill\break
\item[]{\it Catalogue number\/}:~..........\hfill\break
\item[]{\it Program obtainable from\/}:~   \hfill\break
\item[]{\it Computer for which the program is designed and others 
on which it has been tested\/}:~any Unix work-station. \hfill\break
\item[]{\it Operating system\/}:~Unix \hfill\break
\item[]{\it Programming language used\/}:~C++ \hfill\break
\item[]{\it No. of lines in combined program and test 
deck\/}:                                       \hfill\break
\item[]{\it Keywords\/}:~ B-splines, Finite Element Method,
Lagrange type shape functions, relativistic mean-field theory, 
mean-field approximation, spherical nuclei,  
Dirac equations, Klein-Gordon equations, 
classes \hfill\break \vskip 0.5cm
\item[]{\it Nature of physical problem}\hfill\break
The ground-state of a spherical nucleus is 
described in the framework of relativistic mean field 
theory in coordinate space. The model describes a
nucleus as a relativistic system of baryons and mesons. 
Nucleons interact in a relativistic covariant manner 
through the exchange of virtual mesons:~the isoscalar scalar $\sigma$-meson, 
the isoscalar vector $\omega$-meson and the isovector vector $\rho$-meson. 
The model is based on the one boson exchange description of the 
nucleon-nucleon 
interaction.
\\
\item[]{\it Method of solution}\hfill\break
An atomic nucleus is described by a coupled system of partial
differential equations for the nucleons (Dirac equations),
and differential equations for the meson and photon
fields (Klein-Gordon equations). Two methods are compared which allow
a simple, self-consistent solution based on finite element analysis.
Using a formulation based on weighted residuals, the coupled system
of Dirac and Klein-Gordon equations is transformed into a generalized
algebraic eigenvalue problem, and systems of linear and nonlinear  
algebraic equations, respectively. 
Finite elements of arbitrary order are used on uniform radial mesh.
B-splines are used as shape functions in the finite elements.
The generalized eigenvalue problem is solved in narrow windows of the
eigenparameter using a highly efficient bisection method for band matrices.
A biconjugate gradient method is used for the solution of systems of linear
and nonlinear algebraic equations. 

\item[]{\it Restrictions on the complexity of the problem}\newline
In the present version of the code we only consider nuclear systems 
with spherical symmetry. 

\end{itemize}
\bigskip
\centerline{\large \bf LONG WRITE-UP}
%
\section {Introduction}
\noindent 
Over the last decade, the relativistic mean field theory (RMF) has been
applied with great success to the description of low energy properties of
nuclei \cite{Rin.96,Rei.89} and to the description of scattering
at intermediate energies \cite{}. Therefore, RMF gains increasing
recognition. Effective models have been suggested \cite{GRT.89,Rei.89}
which are represented by Lagrangians containing both, nucleonic and
mesonic fields with coupling constants that have been
adjusted to the many body system of nuclear matter and to finite nuclei in
the valley of $\beta$-stability \cite{Lala.96,RRM.86}. Of course, such
a procedure is completely phenomenological and in spirit very similar
to the non-relativistic density dependent HF-models (DDHF) of
Skyrme and Gogny \cite{Sky.72,Gog.80}. Compared to DDHF theory, the
relativistic models seem to have important advantages:
(i) they start on a more fundamental level, treating mesonic degrees 
explicitly and allowing a natural extension for heavy-ion reactions with
higher energies, (ii) they incorporate from the beginning important 
relativistic effects, such as the existence of two types of potentials
(scalar and vector) and the resulting strong spin-orbit term, a new
saturation mechanism by the relativistic quenching of the attractive
scalar field and the existence of anti-particle solutions,
(iii) finally they are in many respects easier to handle than
non-relativistic DDHF calculations. \newline

Since the discovery of the halo phenomenon in light drip-line nuclei
\cite{Tani.85} the study of the structure of exotic nuclei has become
a very exciting topic. Experiments with radioactive beams provide a lot
of new data over entirely new (''exotic'') regions of the chart of 
nuclides. On the theoretical side, presently existing models of the
nucleus, relativistic ones as well as non-relativistic ones, have to be
tested in these new regions in comparison with experiment.
Improvements and extensions of the models become necessary.

Recent investigations \cite{PVLR.97,DFT.84} have shown that coupling
to the particle continuum and large extensions in coordinate space have
to be taken into account in order to describe phenomena of exotic
nuclear structure. The underlying equations of all nuclear models have
therefore to be solved on discretizations in coordinate space. In
contrast to ''conventional'' methods, based on expansions of the
solution in basis functions with spherical or axial symmetry, sophisticated
techniques have to be applied in order to solve the mean field equations
in coordinate space.

With the non-relativistic HF-models extensive nuclear structure calculations
have been performed based on the imaginary time method \cite{Floc.80}.
This very efficient method, however, is restricted to the non-relativistic
cases where the single particle spectrum is limited from below.
In relativistic model calculations, the imaginary time method would
not converge due to mixing with negative energy states. Therefore, we plan
a different approach with Krylov-subspace based methods \cite{Marq.95} 
(for solutions on 2D and 3D meshes in coordinate space) 
and with the bisection method (1D spherical case \cite{PVRR.97}).
In contrast to the imaginary time method, the required single particle
or quasi particle eigenstates have to be calculated in each step of a
self-consistent iteration. At first sight, it seems, that this 
approach is intractable
since coordinate space discretizations on 2D or 3D finite element meshes
lead to eigenvalue problems of large dimensions. With the block Lanczos
method however, the calculation of eigenvalues and corresponding
eigenvectors can be restricted to a small number which is required in
the region of bound nucleons. In combination with the selfconsistent
iteration method which is applied to the whole problem, 
the number of internal block Lanczos
iterations can be reduced to corrections of the vectors which come 
from the previous iterations step of the selfconsistent loop. 
In references \cite{PVR.96,PVRR.97} 
the solution of the spherical RMF equations and the spherical RHB equations
with the finite element method in coordinate space
has been demonstrated. In these
investigations, I have observed that spurious solutions appear in the
spectrum of eigenvalues of the Dirac operator of the RHB equations 
when they are discretized with finite elements of the Lagrangian type. 
Since the numerical
cost to calculate eigensolutions on 1D-meshes is relatively small,
it was not important to avoid spurious solutions a priori and therefore 
they have been eliminated by comparison of the number of nodes.
In the 2D and 3D cases, however, it is important to reduce the size of
the stiffness matrices to a minimum. This can be achieved by using 
shape functions with extremely good properties of interpolation, allowing
wider meshes in coordinate space. Since B-splines are smooth, one would
expect that they have the desired properties.

The major goal of the present paper is to give an answer to the
question whether B-splines can improve the numerics in comparison
to the often used shape functions of Lagrangian type. At the present
state, our study is restricted to the solution of relativistic 
mean field equations. The results of our investigation are important
with respect to large scale computations on finite element meshes of
two and three dimensions. Such calculations are required in the 
relativistic mean field description of deformed exotic nuclei at low
energies.
I have worked out a B-spline version of the computer code which is
published in \cite{PVRR.97} and compare the results obtained with
both codes for spherical nuclei.
\vskip 0.5 cm
%
%
%
%
%
\section {The relativistic mean field equations}
%
%
%
The relativistic mean field model describes the nucleus as a 
system of nucleons which interact
through the exchange of virtual mesons:~the isoscalar scalar $\sigma$-meson, 
the isoscalar vector $\omega$-meson and the isovector vector $\rho$-meson. 
The model is based on the one boson exchange description of the 
nucleon-nucleon interaction. The 
effective Lagrangian density is~\cite{GRT.89}
\begin{eqnarray}
{\cal L}&=&\bar\psi\left(i\gamma\cdot\partial-m\right)\psi
\nonumber\\
&&+\frac{1}{2}(\partial\sigma)^2-U(\sigma )
-\frac{1}{4}\Omega_{\mu\nu}\Omega^{\mu\nu}
+\frac{1}{2}m^2_\omega\omega^2
-\frac{1}{4}{\vec{\rm R}}_{\mu\nu}{\vec{\rm R}}^{\mu\nu}
+\frac{1}{2}m^2_\rho\vec\rho^{\,2}
-\frac{1}{4}{\rm F}_{\mu\nu}{\rm F}^{\mu\nu}
\nonumber\\
&&-g_\sigma\bar\psi\sigma\psi-
g_\omega\bar\psi\gamma\cdot\omega\psi-
g_\rho\bar\psi\gamma\cdot\vec\rho\vec\tau\psi -
e\bar\psi\gamma\cdot A \frac{(1-\tau_3)}{2}\psi\;.
\label{Lagrangian}
\end{eqnarray}
Vectors in isospin space are denoted by arrows.
The Dirac spinor $\psi$ denotes the nucleon with mass $m$.
$m_\sigma$, $m_\omega$, and $m_\rho$ are the masses of the
$\sigma$-meson, the $\omega$-meson, and the $\rho$-meson.
$g_\sigma$, $g_\omega$, and $g_\rho$ are the
corresponding coupling constants for the mesons to the
nucleon. $e^2 /4 \pi = 1/137.036$.

Since the relativistic mean field model has been described in a large
number of articles, I omit a long discussion of the above given Lagrangian
and the derivation of the RMF equations. Instead, I refer to section 2
of reference \cite{PVR.96} and to section 2 of reference \cite{PVRR.97}.
In these references, the development preceding to the investigations
of the present paper is described in details. The main interest of the
work presented below is focused on numerical aspects and performance
of two FEM techniques in the solution of the RMF equations for spherical
nuclei. In the following, I briefly list the static RMF equations for
the spherical symmetric case.

Introducing spherical polar coordinates ($r,\theta,\phi$), the
Dirac equation reduces to a set of two coupled ordinary differential
equations for the amplitudes $g(r)$ and $f(r)$ for proton and neutron
states
\begin{eqnarray}
\label{Equ.2.1}
\Biggl(\partial_r+{{\kappa +1}\over{r}}\Biggr)\,g(r) +
\bigl( m+S(r)-V(r)\bigr)\,f(r)&=&-\varepsilon\, f_i(r),    \,\nonumber \\
\Biggl(\partial_r-{{\kappa -1}\over{r}}\Biggr)\,f(r) +
\bigl( m+S(r)+V(r)\bigr)\,g(r)&=&+\varepsilon\, g(r),
\end{eqnarray}
where the quantum number $\kappa = \pm 1,\pm 2, \pm 3,...$.
The scalar potential $S(r)$ and the vector potential $V(r)$ are composed
of boson field amplitudes and coupling constants where
\begin{equation}
\label{Equ.2.2}
S(r)\,=\,g_{\sigma}\,\sigma(r),
\end{equation}
and 
\begin{equation}
\label{Equ.2.3}
V(r)=g_{\omega}\,\omega^0(r)+
g_{\rho}\,\tau_3\,\rho^0_3(r)
+e{{(1-\tau_3)}\over 2}A^0(r).
\end{equation}
The symbols $g_{\sigma}$, $g_{\omega}$, $g_{\rho}$, and $e$ denote the
coupling constants of the $\sigma$-field, the $\omega$-field, the 
$\rho$-field and the $A$-field, coupled to the nucleons.
The meson fields $\sigma(r)$, $\omega^0(r)$, $\rho^0_3(r)$ and the
photon field $A^0(r)$ are solutions of the inhomogeneous Klein-Gordon
equations
\begin{eqnarray}
\label{Equ.2.4.a}
\Bigl( -\partial_r^2-{2\over r}\partial_r+{{l(l+1)}\over{r^2}}+
m_{\sigma}^2 \Bigr)\,\sigma(r)\, &=&
-g_{\sigma}\,\rho_{\rm s}(r)-g_2\,\sigma^2(r)-g_3\,\sigma^3(r) \, \\
\label{Equ.2.4.b}
\Bigl( -\partial_r^2-{2\over r}\partial_r+{{l(l+1)}\over{r^2}}+
m_{\omega}^2\Bigr)\,\omega^0(r) &=&
g_{\omega}\,\rho_{\rm v}(r) \, \\
\label{Equ.2.4.c}
\Bigl( -\partial_r^2-{2\over r}\partial_r+{{l(l+1)}\over{r^2}}+
m_{\rho}^2\Bigr)\,\rho^0(r) &=&
g_{\rho}\,\rho_3(r) \, \\
\label{Equ.2.4.d}
\Bigl( -\partial_r^2-{2\over r}\partial_r+{{l(l+1)}\over{r^2}}
\Bigr)\,A^0(r) &=&
e\,\rho_{\rm em}(r)
\end{eqnarray}
where the sources of the fields are the scalar density $\rho_s(r)$,
the isoscalar baryon density $\rho_v(r)$, the isovector baryon density
$\rho_3(r)$ and the electromagnetic charge density. They are composed
of the nucleon wave functions in a bilinear way as
\begin{eqnarray}
\label{Equ.2.5.a}
\rho_{\rm s}(r) &=&\sum\limits_{\kappa,\,n} n_{\kappa,n}
{{(2\vert\kappa\vert}\over{4\pi}}
\Bigl(g_{\kappa,n}(r)^2 - f_{\kappa,n}(r)^2\Bigr) \, \\
\label{Equ.2.5.b}
\rho_{\rm v}&=&\sum\limits_{\kappa,n} n_{\kappa,n}
{{2\vert\kappa\vert}\over{4\pi}}
\Bigl(g_{\kappa,n}(r)^2 + f_{\kappa,n}(r)^2\Bigr) \, \\
\label{Equ.2.5.c}
\rho_{3}&=&\sum\limits_{\kappa,n} n_{\kappa,n} \tau_{3n}
{{2\vert\kappa\vert}\over{4\pi}}
\Bigl( g_{\kappa,n}(r)^2+\vert f_{\kappa,n}(r)^2\Bigr) \, \\
\label{Equ.2.5.d}
\rho_{\rm em}&=&\sum\limits_{\kappa,n} n_{\kappa,n}{{(1-\tau_{3n})}\over2}
{{2\vert\kappa\vert}\over{4\pi}}
\Bigl(g_{\kappa,n}(r)^2+ f_{\kappa,n}(r)^2 \Bigr) 
\end{eqnarray}
where the quantities $n_{\kappa,n}$ are occupation numbers of the
energy levels (indices $\kappa$,$n$). For the simple Hartree model
without pairing, $n_{\kappa,n}=1$ for occupied levels and $n_{\kappa,n}=0$
for unoccupied levels. The index $n$ denotes the principal quantum number
($n=0,1,2,...$) and counts the eigensolutions of equation (\ref{Equ.2.1})
from small to large energies $\varepsilon_{\kappa,n}$. The nucleon numbers
filling an orbital $(\kappa,n)$ are taken into account by the factors
$2\vert\kappa\vert$ in Eqs. (\ref{Equ.2.5.a}) - (\ref{Equ.2.5.d}).
Since the densities (\ref{Equ.2.5.a}) - (\ref{Equ.2.5.d}) do not
depend on the angular coordinates $\theta$ and $\phi$, no terms higher
than of monopole order show up at the r.h.s. of 
Eqs. (\ref{Equ.2.4.a}) - (\ref{Equ.2.4.d}). Consequently, the solution
of the these equations has to be restricted to $l=0$ in the description of
spherical nuclei. For physical reasons, the nonlinear self-coupling of the
$\sigma$-field has to be taken into account. It is described by the two
terms $-g_2\,\sigma^2(r)$ and $-g_3\,\sigma^3(r)$ at the r.h.s. of
Eq. (\ref{Equ.2.4.a}). Without these terms, the RMF-model could not 
explain the compressibilities in finite nuclei and nuclear matter as
well as the surface properties of finite nuclei.
\vskip 2.0cm
%
%
%
%
\section {B-spline and Langrangian type finite elements}
%
\smallskip
The most widely used finite element type in many applications is the
Lagrange type element. Lagrangian shape functions allow the
simplest representation compared to other types of shape functions.
For any finite element order $n$ they have the following 
expression in reference element representation (see Figs. 2a, 2c, 2e)
\begin{equation}
N_q^n(\rho)=\prod^n_{l=0 \atop l\ne q} {{(n\rho-l)}\over{(q-l)}}
\label{Equ.3.1}
\end{equation}
where the coordinate $\rho$ is restricted to the interval
$\bigl[ 0, 1\bigr]$. From Eq. (\ref{Equ.3.1}) it becomes obvious that
Lagrange type finite elements are easy to handle. Generally, for any
conventional finite element type, the shape functions (nodal basis) have the
property
\begin{equation}
N_q^n({q'}/n) = \delta_{qq'}
\label{Equ.3.2}
\end{equation}
in the one dimensional case and
\begin{equation}
N_q^n(\vec\rho_{q'}) = \delta_{qq'},\qquad \vec\rho=(\rho_1,...,\rho_M)^T
\label{Equ.3.3}
\end{equation}
in the M-dimensional case where $\vec\rho_{q'}$ denotes a grid point of
M-dimensional finite elements. These shape functions form a so called nodal
basis. Shape functions with property (\ref{Equ.3.3}) can be constructed
for finite elements of any geometrical form as for example triangular
elements or quadratic elements in two dimensions and tetraedric or
cubic elements in three dimensions. Also the location of the mesh points
which belong to a finite element can be distributed in almost arbitrary
manner over the domain of the element. In most cases, however, M-cube
elements (intervals, squares, cubes, etc.) with a uniform distribution of
the nodes are sufficient and allow extremely efficient calculations of the
stiffness matrices for a given boundary value problem. The shape functions
of such elements are represented as products of Lagrange polynomials
(\ref{Equ.3.1})
\begin{equation}
N_{(q_1,...,q_M)}^{(q_1,...,q_M)}(\rho_1,...,\rho_M)=
\prod_{i=1}^M N_{q_i}^{n_i}(\rho_i),\qquad q_i,...,n_i,
\label{Equ.3.4}
\end{equation}
where $n_i$ denotes the order of the element in the direction of
dimension $i$ and $(q)=(q_1,...,q_M)$ forms the index tuple of the nodes.

The construction (\ref{Equ.3.4}) of Lagrange type M-cube shape functions
shows that 1-dimensional Lagrange type finite elements allow the most
simple generalization to M-cube meshes. A great advantage of such
shape functions can also be seen from a technical view point. 
Implementations of the general M-dimensional case in object oriented
programming styles become simple. The amount of data required by an
object which represents a Lagrangian M-cube finite element as reference
element is almost the same as in the 1-dimensional case if the 
orders $n_1,...,n_M$ are equal. In this special case the data 
representing all shape functions of a M-cube element comprise
$2\cdot n_i\cdot n_i^G$ floating point numbers where I denote by $n_i^G$ the
number of Gauss points on a Gaussian mesh in dimension $i$.
In the more general case where $n_i\ne n_j$ for $i\ne j$ and
$n_i^G\ne n_j^G$ for $i\ne j$ the amount of float point numbers required
to represent all shape functions is
\begin{equation}
2\sum_{i=1}^M n_i\, n_i^G
\label{Equ.3.5}
\end{equation}
which is still small compared to the number of values
\begin{equation}
2\prod_{i=1}^M n_i\, n_i^G
\label{Equ.3.6}
\end{equation}
required for shape functions of arbitrary type and arbitrary distribution
of the nodes over the element.

The numerical cost for the integration of matrix elements reduces 
dramatically in cases where operators split up into products of operators
each depending on a complementary subset of the coordinates 
$\rho_1,...,\rho_M$. In the most ideal case an operator factorizes
completely leading with Eq. (\ref{Equ.3.4}) to a complete factorization
of matrix elements.
\begin{equation}
\Bigl< N_{(q_1,...,q_M)}^{(n_1,...,n_M)}(\rho_1,...,\rho_M)\Big\vert
\prod_{i=1}^M \hat O_i(\rho_i)\Big\vert 
N_{(q_1',...,q_M')}^{(n_1,...,n_M)}(\rho_1,...,\rho_M)\Bigr>
=\prod_{i=1}^M\Bigl< N_{q_i}^{n_i}(\rho_i)\Big\vert\hat O_i(\rho_i)
\Big\vert N_{q_i'}^{n_i}(\rho_i)\Bigr>.
\label{Equ.3.7}
\end{equation}
This becomes obvious when I rewrite Eq. (\ref{Equ.3.5}) in terms of a
numerical Gauss integration
\begin{eqnarray}
\sum_{l_1,...,l_M}^N N_{(q)}^{(n)}(\rho_1^{l_1},...,\rho_M^{l_M})
\hat O(\rho_1^{l_1},...,\rho_M^{l_M}) 
N_{(q)}^{(n)}(\rho_1^{l_1},...,\rho_M^{l_M})\prod_{i=1}^M w_{l_i} =
\prod_{i=1}^M\sum_{l_i=1}^N N_{q_i}^{n_i}(\rho_i^{l_i})
\hat O_i(\rho_i^{l_i}) N_{q_i'}^{n_i}(\rho_i^{l_i})
\label{Equ.3.8}
\end{eqnarray}
where, assuming that N is the number of Gauss points in each coordinate,
the number of floating point operations on the left hand side is greater
than $4N^M$ while on the r.h.s. it is smaller than 3MN.

%
%
%
%
%
\begin{figure}[H]
\centerline{
{ \epsfysize=12cm \epsfxsize=16cm                  
\epsffile{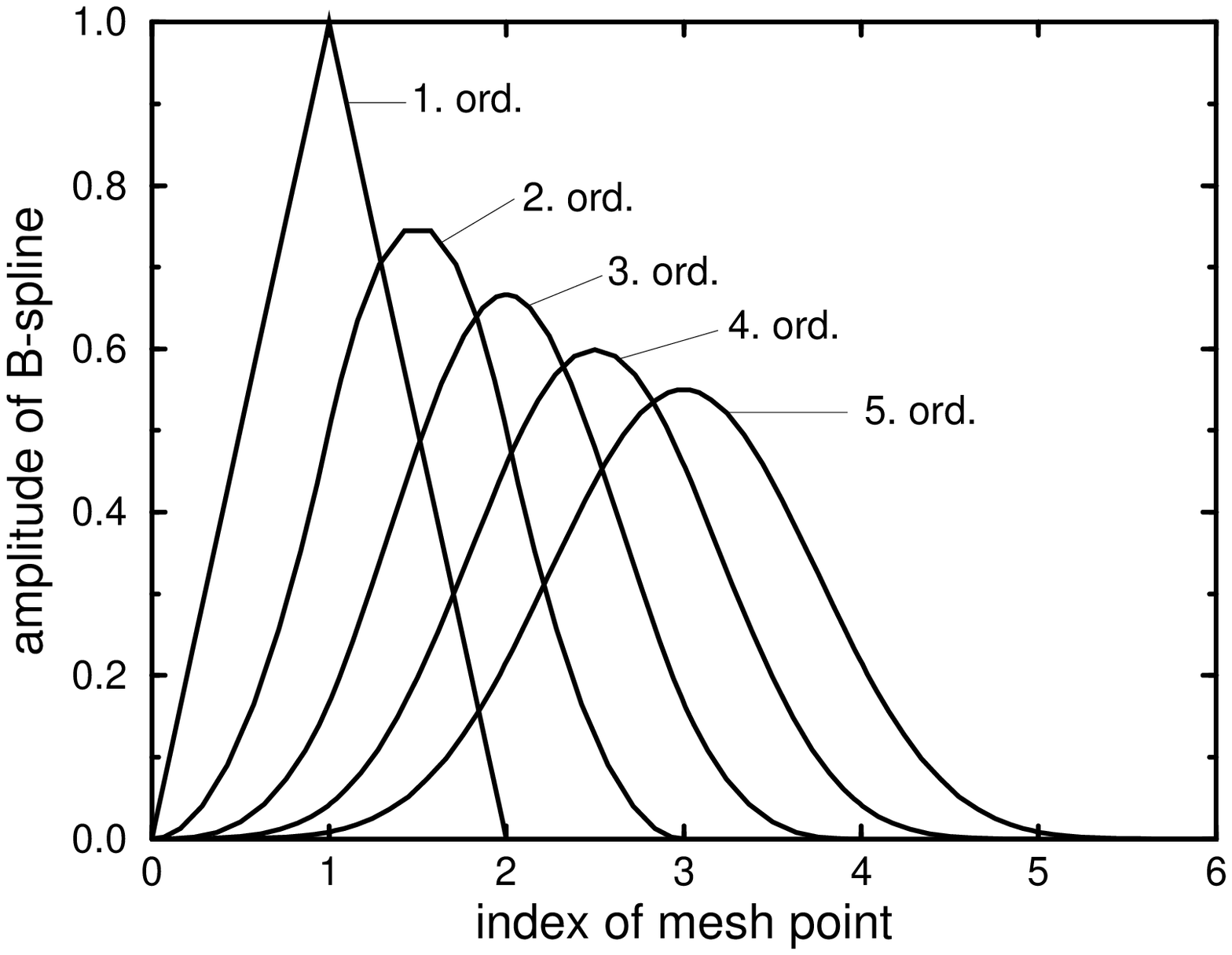} }
}
\vskip 0.2cm    
{\small {\bf Fig. 1:} \quad
B-spline functions of different orders increasing from 1 to 5.
}
\end{figure}
\vskip 0.2cm    
These advantages of Lagrangian type finite elements gave us a reason to
apply them in several previous studies and calculations (see references
\cite{PVR.96,PVRR.97,PVR.97,PVLR.97}). In these references, it has been shown
that Lagrangian finite elements provide an excellent tool for solving the
equations of the RMF model in self-consistent iterations in coordinate space.
In this manuscript, I present a new finite element technique using B-splines as
shape functions and compare this method with the FEM based on Lagrangian
shape functions. B-splines have a compact support and are defined as
polynomials piecewise on intervals which are bounded by neighbored mesh
points. The basic criterium in the construction of these basis functions
is optimized smoothness over the whole support. This property is 
guaranteed if all derivatives up to the order
$n^{}-1$ obey the conditions of continuity in all matching points of the
mesh. The order $n^{}$ of a given B-spline corresponds to the degree
of the polynomials by which it is composed. In Fig. 1, examples are shown for
B-splines of order one to five. $n^{}+2$ mesh points are required to
construct a B-spline of the order $n^{}$. In contrast to Lagrangian shape
functions, B-splines of any order do not change sign. A common property of 
both types of shape functions is expressed by
\begin{equation}
\sum_p N_p(\rho) = \sum_p B_p(\rho) = 1
\label{Equ.3.9}
\end{equation}
where $p$ denotes the mesh point index.
The fact, that B-splines of any order satisfy all these conditions
makes it impossible to find an expression in closed form in the sense
of Eq. (\ref{Equ.3.1}). 
Rather, they are generated by the following brief algorithm.
\begin{eqnarray}
\mbox{start:}\qquad B_{i,1}(x) & = & \left\{
\begin{array}{ll}
{ ({(x_{i+1}-x_i)})^{-1} } & x_i\le x \le x_{i+1}               \cr
  0                     & x < x_i,\, x > x_{i+1} 
\end{array} \right. 
\quad i=0,...,n^{}; 
\nonumber \\
B_{i,k}(x) & = & { {(x-x_i)} \over{(x_{i+k}-x_i)} } B_{i,k-1}(x) +
             { {(x_{i+k}-x)} \over{(x_{i+k}-x_i)} } B_{i+1,k-1}(x)
\label{Equ.3.10}
\end{eqnarray}
I define a B-spline finite element as a region which is bounded by two
neighboring mesh points in the one-dimensional case or as a M-cube where
the $2^M$ corners are identical with the $2^M$ mesh points of a cubic
grid which are closest to the center of the cube. Obviously, this
definition is restricted to cubic grids but it will turn out to be extremely
efficient in all cases where cubic finite element discretizations can
be applied.

The figures 2b, 2d, and 2f display $3^{\rm rd}$ order, $4^{\rm th}$ order and $5^{\rm th}$ order 
B-spline finite elements in one dimension according to our definition. 
The figures show, that all parts of a B-spline are contained in each
element.
%
%
%
%
%
\begin{figure}[H]
\centerline{
{ \epsfysize=5cm \epsfxsize=5cm                  
\epsffile{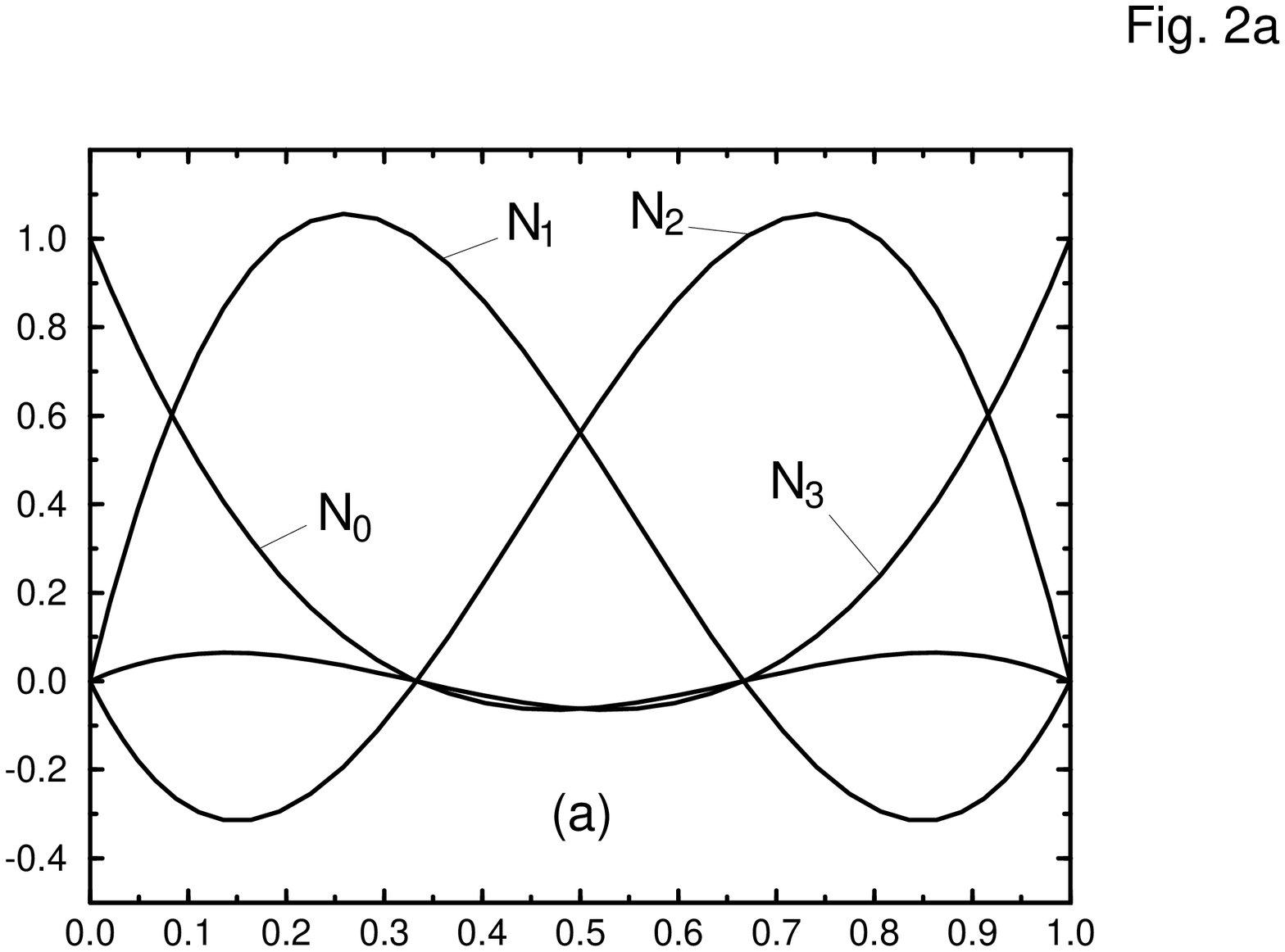} }          \hspace{1cm}
{ \epsfysize=5cm \epsfxsize=5cm                  
\epsffile{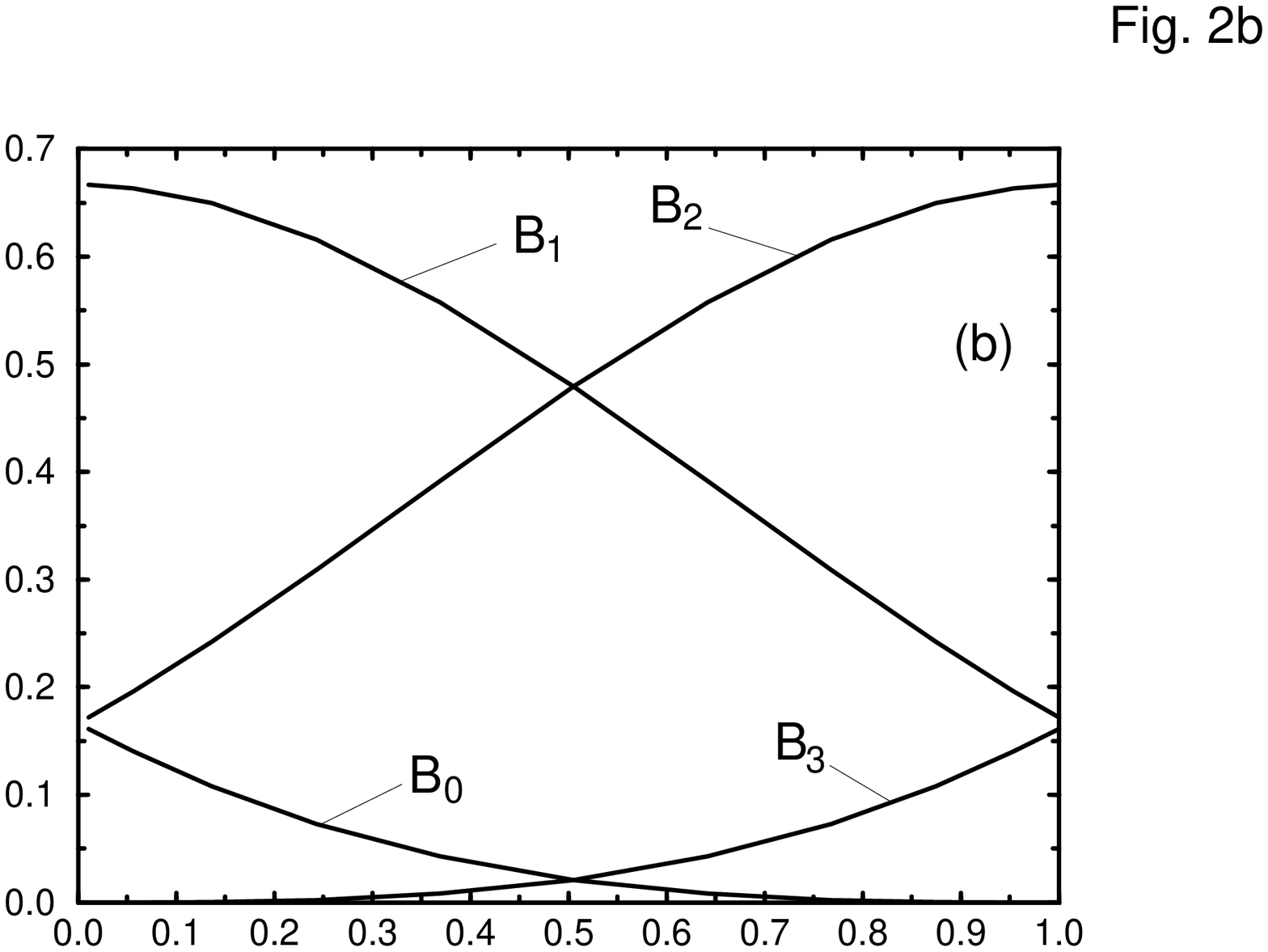} }
}
\end{figure}
\begin{figure}[H]
\centerline{
{ \epsfysize=5cm \epsfxsize=5cm                  
\epsffile{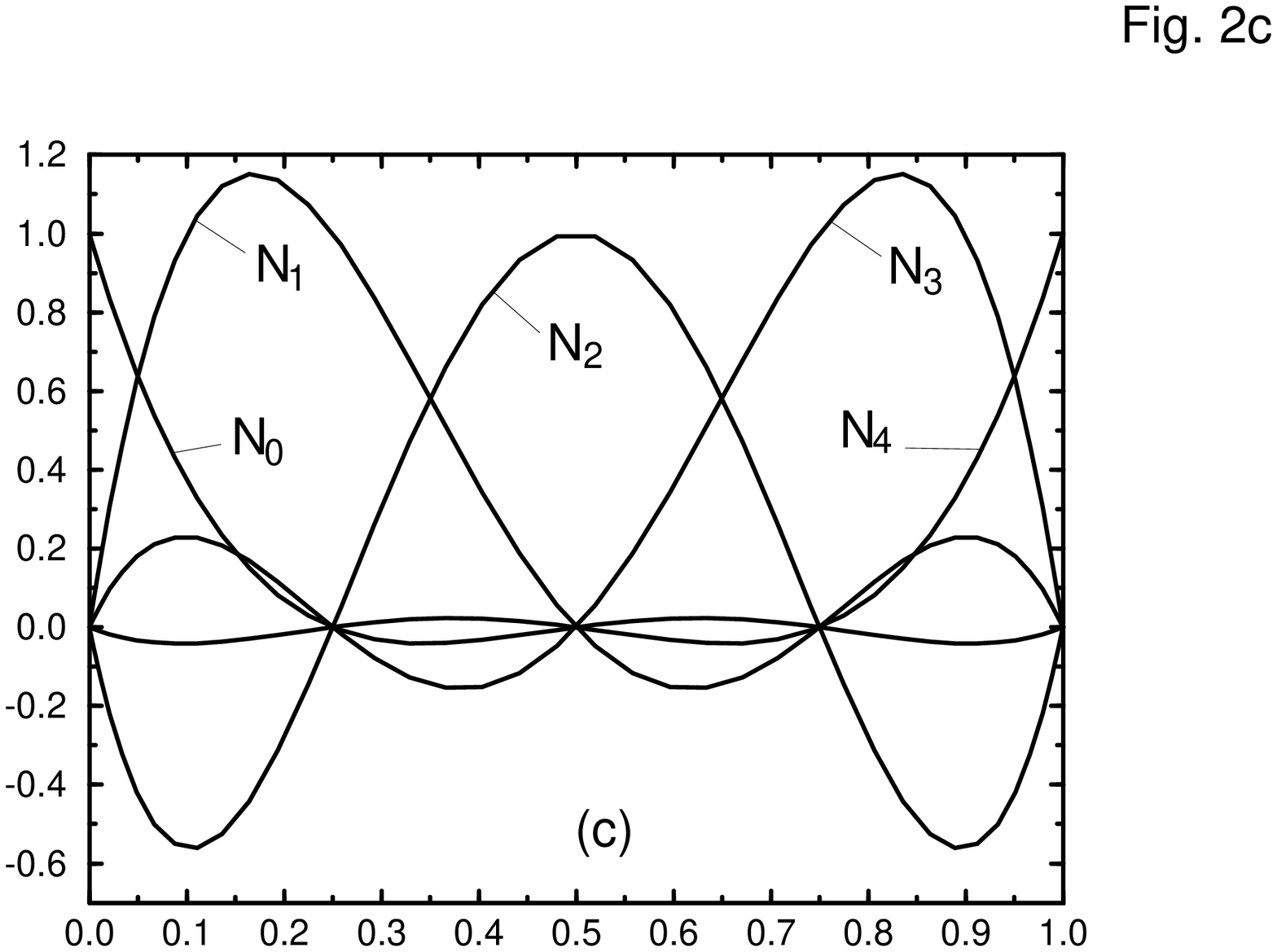} }          \hspace{1cm}
{ \epsfysize=5cm \epsfxsize=5cm                  
\epsffile{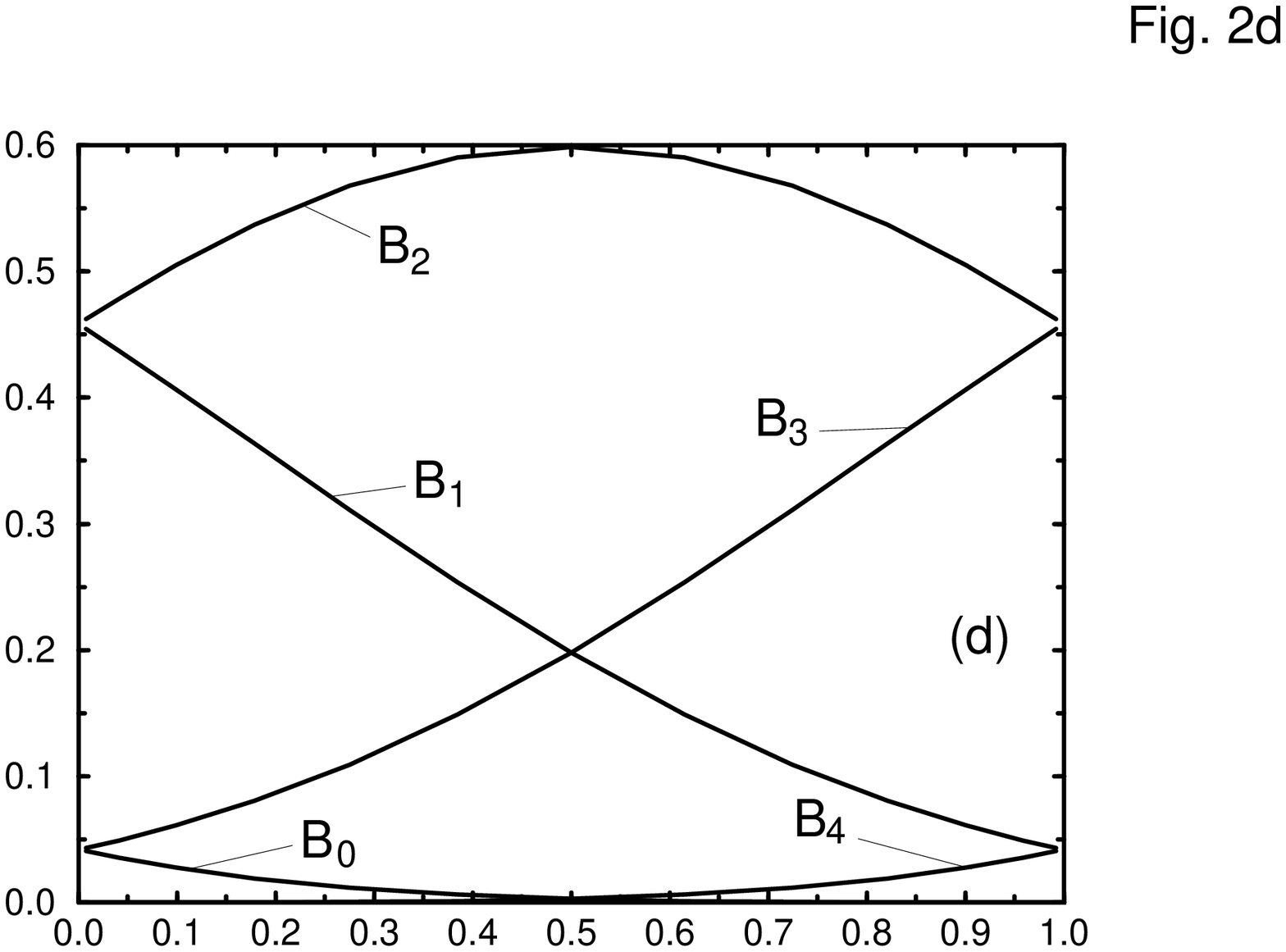} }
}
\end{figure}
\begin{figure}[H]
\centerline{
{ \epsfysize=5cm \epsfxsize=5cm                  
\epsffile{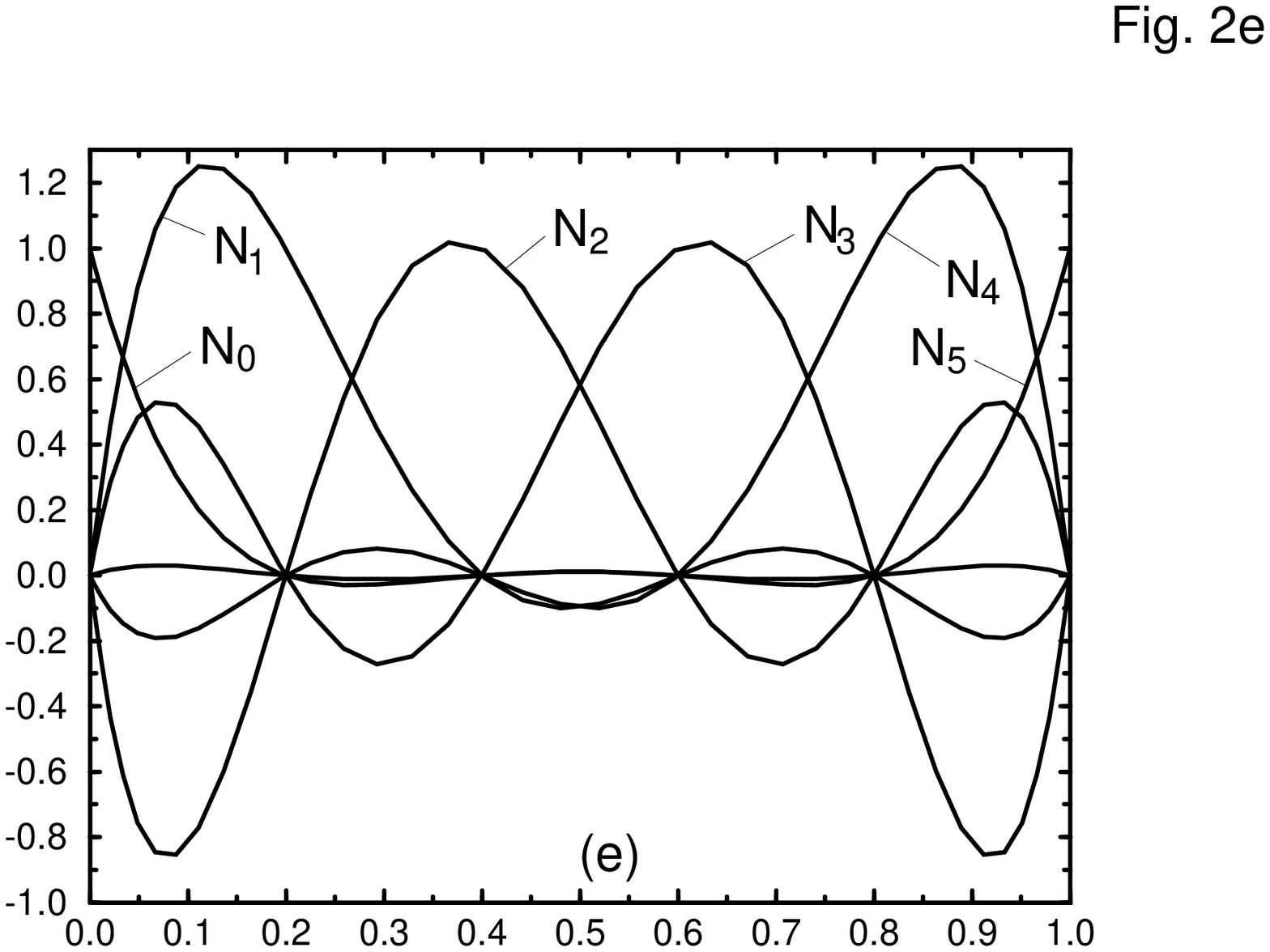} }          \hspace{1cm}
{ \epsfysize=5cm \epsfxsize=5cm
\epsffile{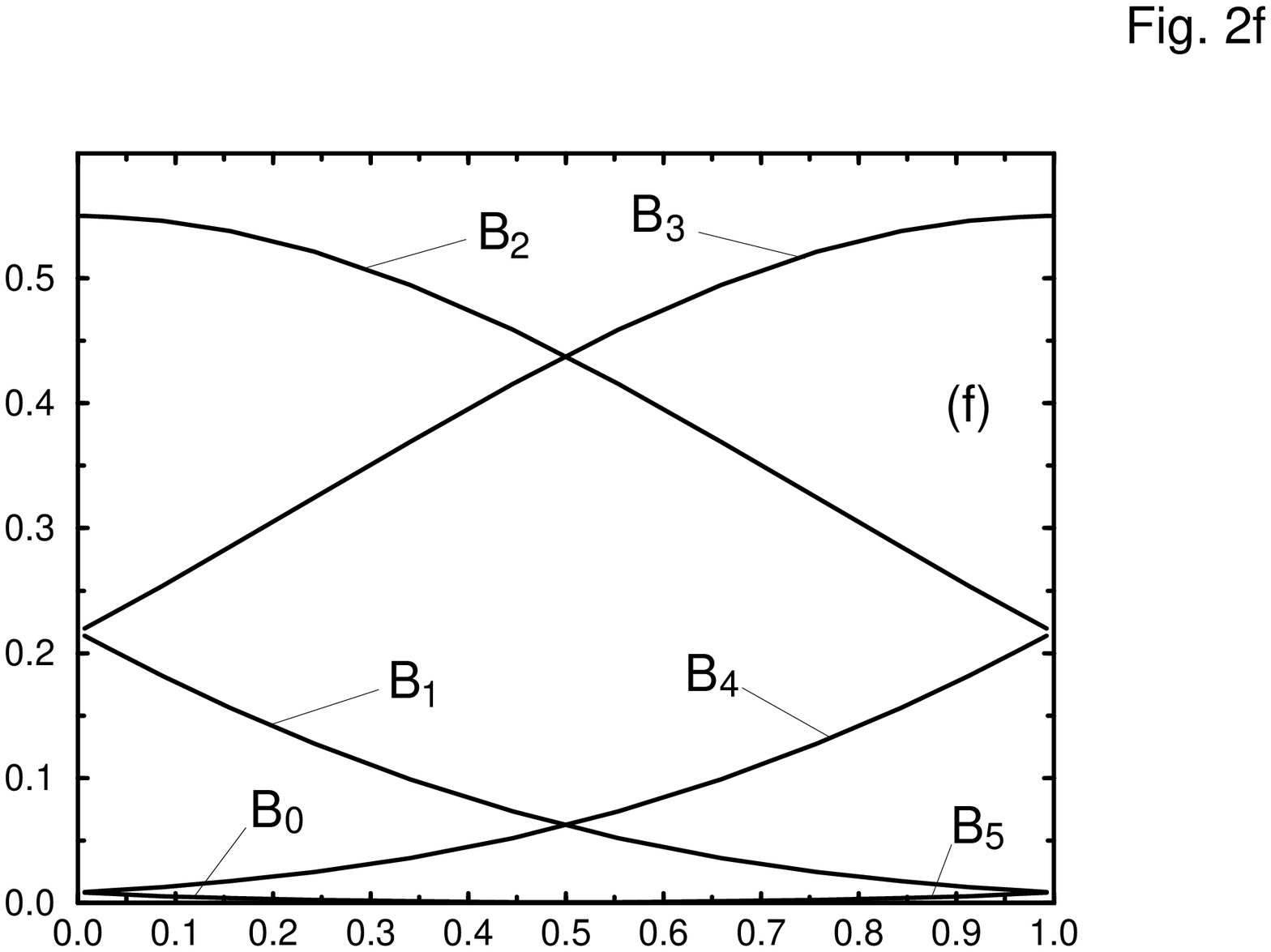} }
}
\vskip 0.5cm    
{\small {\bf Fig. 2:} \quad
Finite elements in reference element representation. In the
figures (a),(c),(e), Lagrange type elements
with corresponding shape functions are shown
for $3^{\rm rd}$ order, $4^{\rm th}$ order and $5^{\rm th}$ order. 
In comparison, B-spline elements
of $3^{\rm rd}$ order, $4^{\rm th}$ order and $5^{\rm th}$ 
order are displayed in the figures (b), (d), (f). 
}
\end{figure}
\vskip 0.5cm    
A B-spline of order $n^{}$ is determined by $n^{}+2$ mesh-points
through the algorithm (\ref{Equ.3.10}) and the support is composed by
$n^{}+1$ intervals.
However, the pieces which are contained in a single element belong to different
B-splines each located at another mesh point and determined by a different
set of mesh points. They are the overlap sections of all splines which are not
zero in the considered element. This fact has to be taken into account
especially if nonuniform meshes are used. As a consequence, the 
amplitudes of the shape functions in such a finite element depend on the
position of all mesh points which are in the support of all contributing
B-splines. Most of these mesh points are outside of the element and
belong to other elements. Therefore, the degree of ''interaction'' between
neighbored elements is maximized and much higher than for elements of
Lagrange type where only next neighbor elements ''interact'' through
their outermost contact grid points. This leads to a strong reduction of the
total number of required mesh points as I will demonstrate in section 5.

A disadvantage, however, appears when non-uniform meshes are used. As it
has been
discussed in reference \cite{PVRR.97}, Lagrangian elements can be mapped
on a reference element through linear affine transformations. This is even
possible for non-uniform meshes. Therefore, Lagrangian shape functions need
to be evaluated only once in the reference element and amplitudes and
abscissas can be accessed by means of a 
pointer to that reference element. These
advantages are also valid for B-splines as long as uniform meshes are used.
In the case of non-uniform meshes the algorithm (\ref{Equ.3.10}) has to be
evaluated for each argument taken on the global region. This leads to a
reduction of storage requirement but increases the numerical cost by a factor
which corresponds to the number of floating point operations which are
necessary to carry out the scheme (\ref{Equ.3.10}). Consequently, the
numerical effort in the calculation of the stiffness matrices of a given 
problem increases by the same factor. On the other hand, the number of 
algebraic equations resulting from a finite element discretization is
usually large and the numerical cost to solve these equations increases
faster with the number of equations ($\sim $ number of mesh points) than
the numerical cost in the calculation of the stiffness matrices with the
number of mesh points. This trend is even enhanced for increasing
dimension $M$ of the descretization where the condition of the stiffness
matrices becomes worse. B-splines finite elements might therefore be superior
in multidimensional FEM discretizations as compared to Lagrangian finite
elements.

\vskip 1.0cm
%
%
%
%
\section {The FEM discretization}
%
\smallskip
A basic principle of FEM is the approximation of the solution for a given
problem in a space of shape functions which have compact support and existing
continuous weak derivative of maximum degree m. Together with a p-norm which
is for all those functions defined as
\begin{eqnarray}
\Vert\varphi\Vert_{m,p} := \Bigl(\sum_{\alpha = 0}^m \int_\Omega 
\vert D^\alpha\varphi(x) \vert^p \Bigr)^{1/p},
\label{Equ.4.1}
\end{eqnarray}
the above given space is a Banach space. According to the norm it is 
usually called Sobolev space $W_p^m(\Omega)$ where $\Omega$ may be any 
compact domain of the coordinate space. Since $W_p^m(\Omega)$ is complete,
the solutions of any partial differential equation of an order not higher
than m can be approximated to arbitrary precision in $W_p^m(\Omega)$ on
the whole domain $\Omega $. This property plays an important role for the 
solution of differential equations with computational methods because finite
element discretizations of the domain $\Omega $ correspond to subsets of
linear independent functions of $W_p^m(\Omega)$ and because the
representation of functions of $W_p^m(\Omega)$ on the computer is
simple. In FEM, $\Omega $ is subdivided into a large number of small
sub-domains which are called finite elements. Each element is support of a
certain number of shape functions which is equivalent to the number of 
constraints set on the element. These functions span up a finite
element space. The corners of the elements are located on a 
finite element mesh. However, additional mesh points can exist in
the interior or on the surface of each element and additional constraints
as derivatives of any order can be applied. 
In such cases the order of the element is higher than first order.

In the following, I discuss the finite element discretization of the
Eqs. (\ref{Equ.2.1}) and Eqs (\ref{Equ.2.4.a}) to (\ref{Equ.2.4.d}) for
both types Lagrangian and B-spline elements. In the present application
a nodal basis $\{ N_p(r) \}$ is used in the case of Lagrange elements
and a non-nodal basis $\{ B_p(r) \}$ is used in the case of B-spline elements.
Examples for discretizations with 
elements of $3^{\rm rd}$ order are displayed for
both types in Fig. 3a and 3b. Each Lagrangian element in Fig. 3a has two
boundary nodes and two additional nodes in the interior whereas the
B-spline elements in Fig. 3b are free of interior nodes. In the B-spline
method additional nodes are required outside of the region of integration
in order to generate the shape functions which have non-zero overlap with
the inner region.
I use the notation $N_p(\rho )$ and $B_p(\rho)$ ($0\le q\le n^{}$) for
shape functions in reference element representation, and $N_p(r)$ or
$B_p(r)$ ($1\le p \le n^{nod}$) for shape functions $N_p$ and $B_p$ on the
global mesh in the r-coordinate space.
Using the standard representation for the Pauli matrices, the Dirac equation
(\ref{Equ.2.1}) is written in matrix form
\begin{eqnarray}
\Bigl[(\partial_r +r^{-1})\cdot\sigma_3\sigma_1-\kappa r^{-1}\cdot\sigma_1
      +(m+S(r))\cdot\sigma_3 + V(r)\cdot {\bf 1}_2 \Bigr]\,\Phi(r)=
\varepsilon\, {\bf 1}_2\,\Phi(r).
\label{Equ.4.2}
\end{eqnarray}
%
%
%
%
%
\vskip 0.5cm
\begin{figure}[H]
\centerline{
{ \epsfysize=6cm \epsfxsize=8cm                  
\epsffile{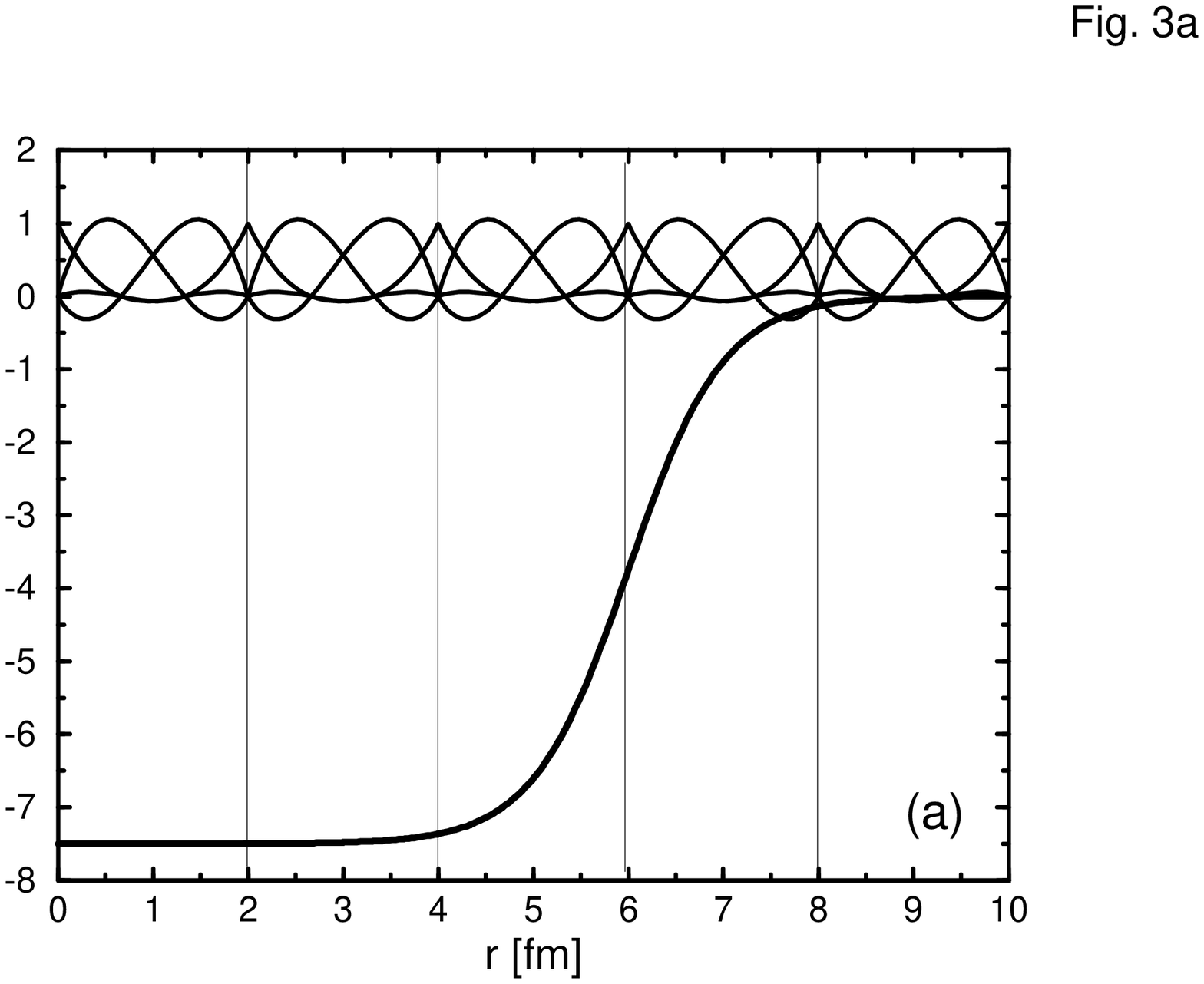} }          \hspace{1cm}
{ \epsfysize=6cm \epsfxsize=8cm                  
\epsffile{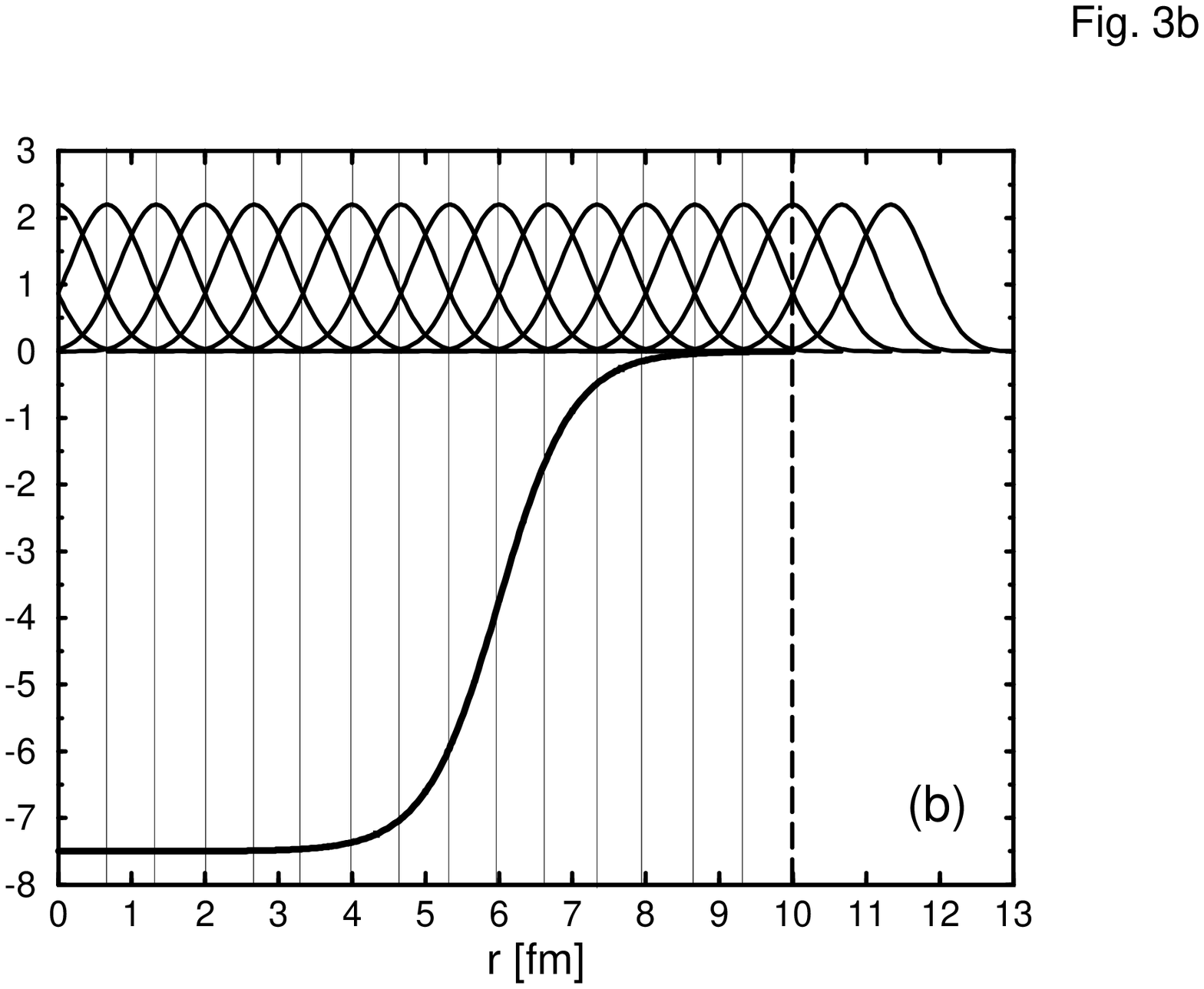} }
}
\vskip 0.5cm
{\small {\bf Fig. 3:} \quad
Global discretizations with 
$3^{\rm rd}$ order finite elements (a) of Lagrangian type
and (b) of B-spline type. In both examples a total number of 16 mesh points
is used in the region between $r=0\,{\rm fm}$ and $r=10\,{\rm fm}$.
}
\end{figure}
\vskip 0.5cm
For the nucleon spinor I use the FEM ansatz
\begin{equation}
\Phi(r) = \sum_p u_p B_p(r)
\label{Equ.4.3}
\end{equation}
with B-spline functions $B_p(r)$ and node variables 
$u_p:= ( u_p^{(g)}, u_p{(f)} )^T $.
In a Lagrangian FEM ansatz, the shape functions $B_p(r)$ defined in 
(\ref{Equ.4.3}) are
replaced by the shape functions $N_p(r)$. In the weak formulation of the
eigenvalue problem (\ref{Equ.2.1}), the weighted residual
(see \cite{PVRR.97}) leads to algebraic equations of the form
\begin{eqnarray}
\sum_p \Bigl< w_{p'}(r)\Big\vert 
\Bigl[(\partial_r +r^{-1})\cdot\sigma_3\sigma_1-\kappa r^{-1}\cdot\sigma_1
      +(m+S(r))\cdot\sigma_3 + V(r)\cdot {\bf 1}_2 \Bigr]\Big\vert
\,B_p(r)\Bigr> u_p = \nonumber \\
\varepsilon\, \Bigl< w_{p'}(r)\Big\vert {\bf 1}_2\Big\vert B_p(r)\Bigr> u_p
\label{Equ.4.4}
\end{eqnarray}
with weighting functions $w_{p'}(r)$. The weighting functions are chosen
\begin{equation}
w_{p'}(r) = \Bigl( 1- \Bigl( {r\over{r_{\rm max}}} \Bigr)^2 \Bigr)\, r^2\,
r^{l_{g/f}}\, B_p(r) 
\label{Equ.4.5}
\end{equation}
in the case of B-spline elements or
\begin{equation}
w_{p'}(r) = \Bigl( 1- \Bigl( {r\over{r_{\rm max}}} \Bigr)^2 \Bigr)\, r^2\,
r^{l_{g/f}}\, N_p(r) 
\label{Equ.4.6}
\end{equation}
when Lagrangian shape functions are used in ansatz (\ref{Equ.4.3}).
\begin{eqnarray}
l_g = \left\{
\matrix{
{ -\kappa - 1}    & \kappa < 0;     \cr
 \, \kappa\qquad  & \kappa > 0;
} \right.
\label{Equ.4.7.a}
\end{eqnarray}
\begin{eqnarray}
l_f = \left\{
\matrix{
{ -\kappa }    & \kappa < 0;     \cr
{\kappa-1 }    & \kappa > 0;
} \right.
\label{Equ.4.7.b}
\end{eqnarray}
The factor $r^2$ corresponds to the Jacobi determinant of the
transformation into spherical coordinates and compensates singularities
in the operator. 
The factor $r^{l_{g}}$ or $r^{l_{f}}$ respectively includes
boundary conditions at $r=0$ properly for upper ($g(r)$) and lower
components ($f(r)$) of the spinor $\Phi(r)$. The factor
$(1-(r/r_{\rm max})^2)$ includes boundary conditions $g(r_{\rm max})=0$ at
$r=r_{\rm max}$ in matrix elements which are multiplied with g-components in 
$\Phi(r)$ whereas this factor is replaced by $1$ when a matrix element is
multiplied with node variables of the $f(r)$-component.  
%
%
%
%
%
\begin{figure}[H]
\centerline{
{ \epsfysize=6cm \epsfxsize=7cm                  
\epsffile{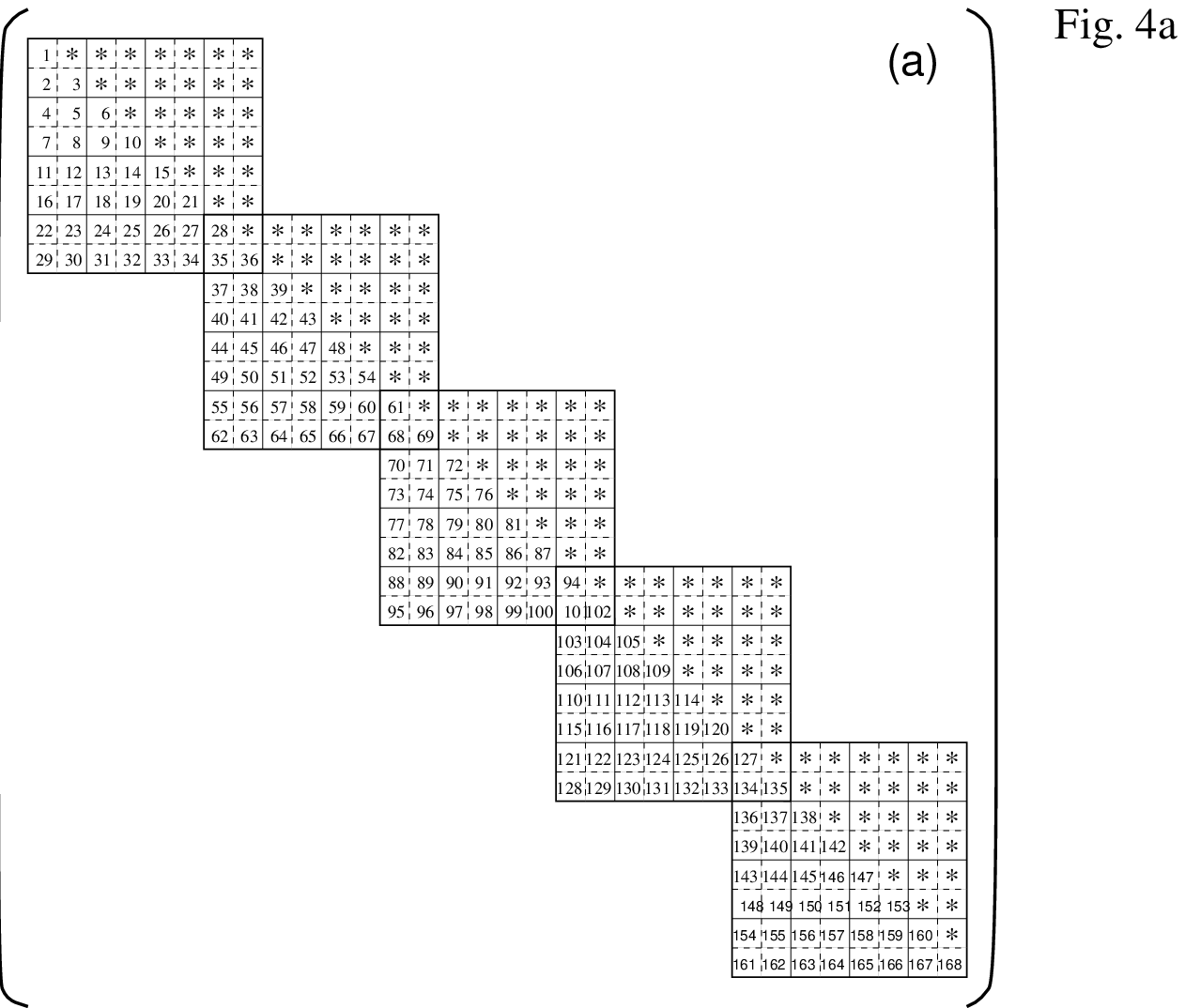} }          \hspace{2cm}
{ \epsfysize=6cm \epsfxsize=7cm                  
\epsffile{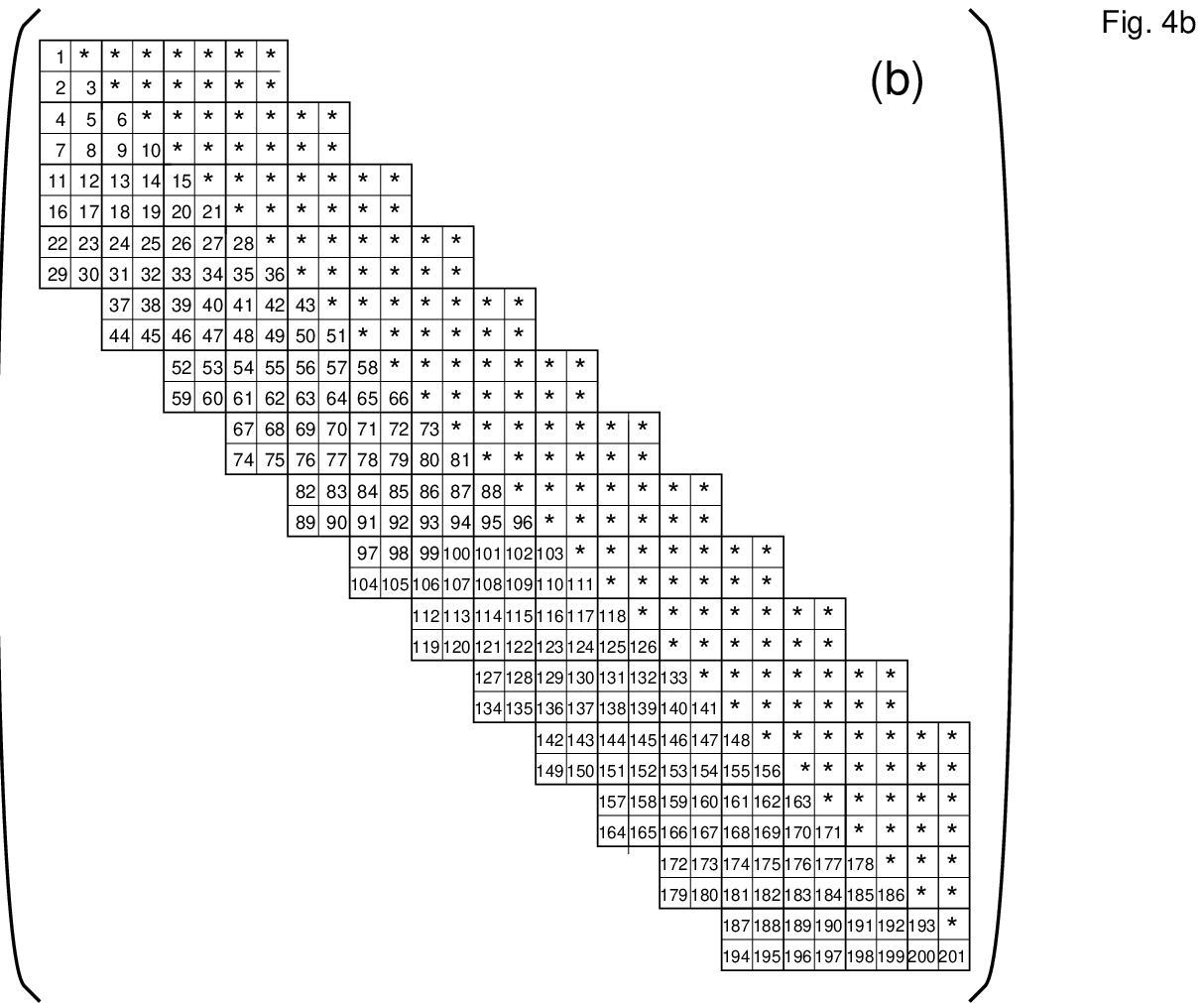} }
}
\end{figure}
\begin{figure}[H]
\centerline{
{ \epsfysize=6cm \epsfxsize=7cm                  
\epsffile{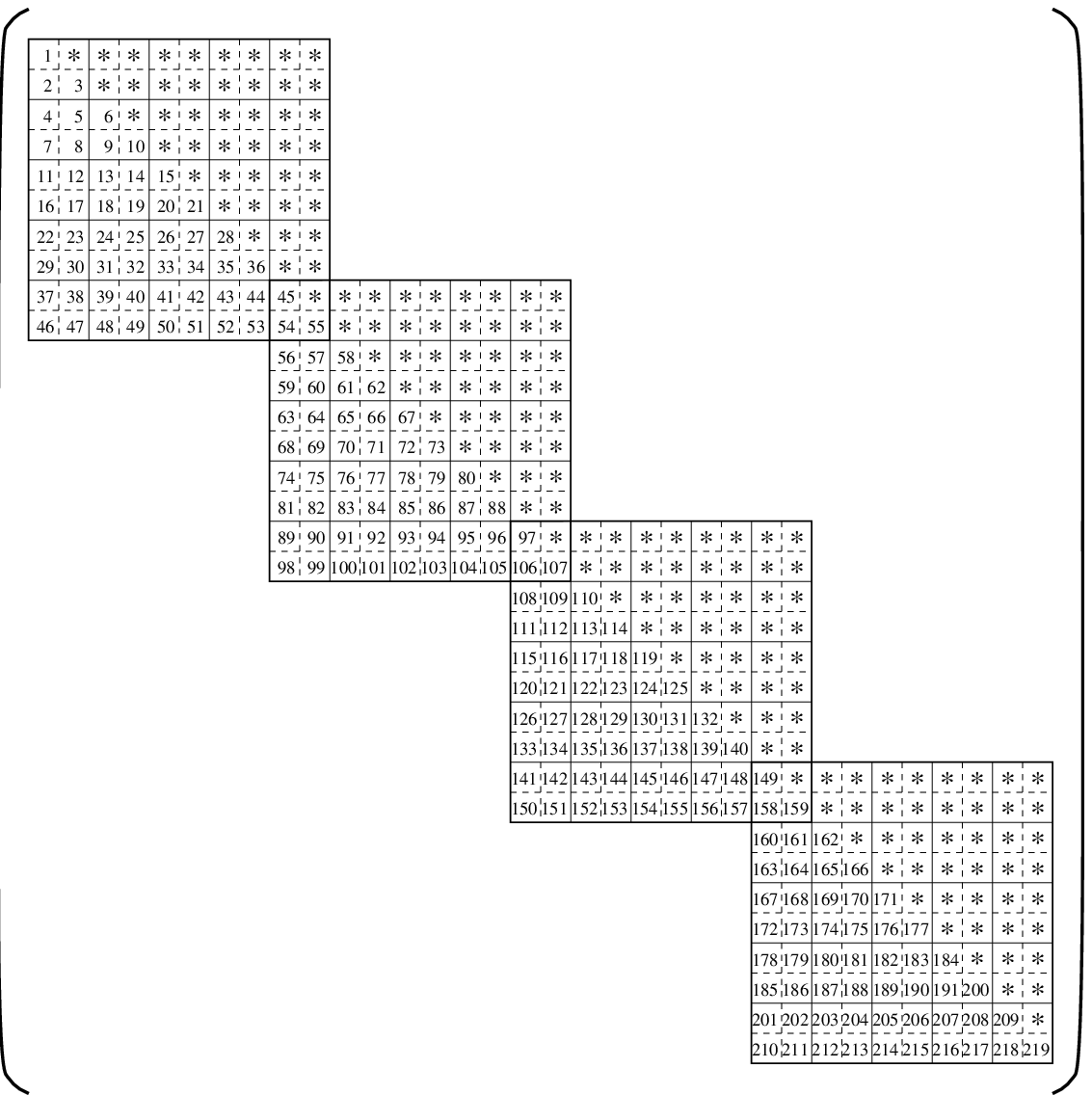} }          \hspace{2cm}
{ \epsfysize=6cm \epsfxsize=7cm                  
\epsffile{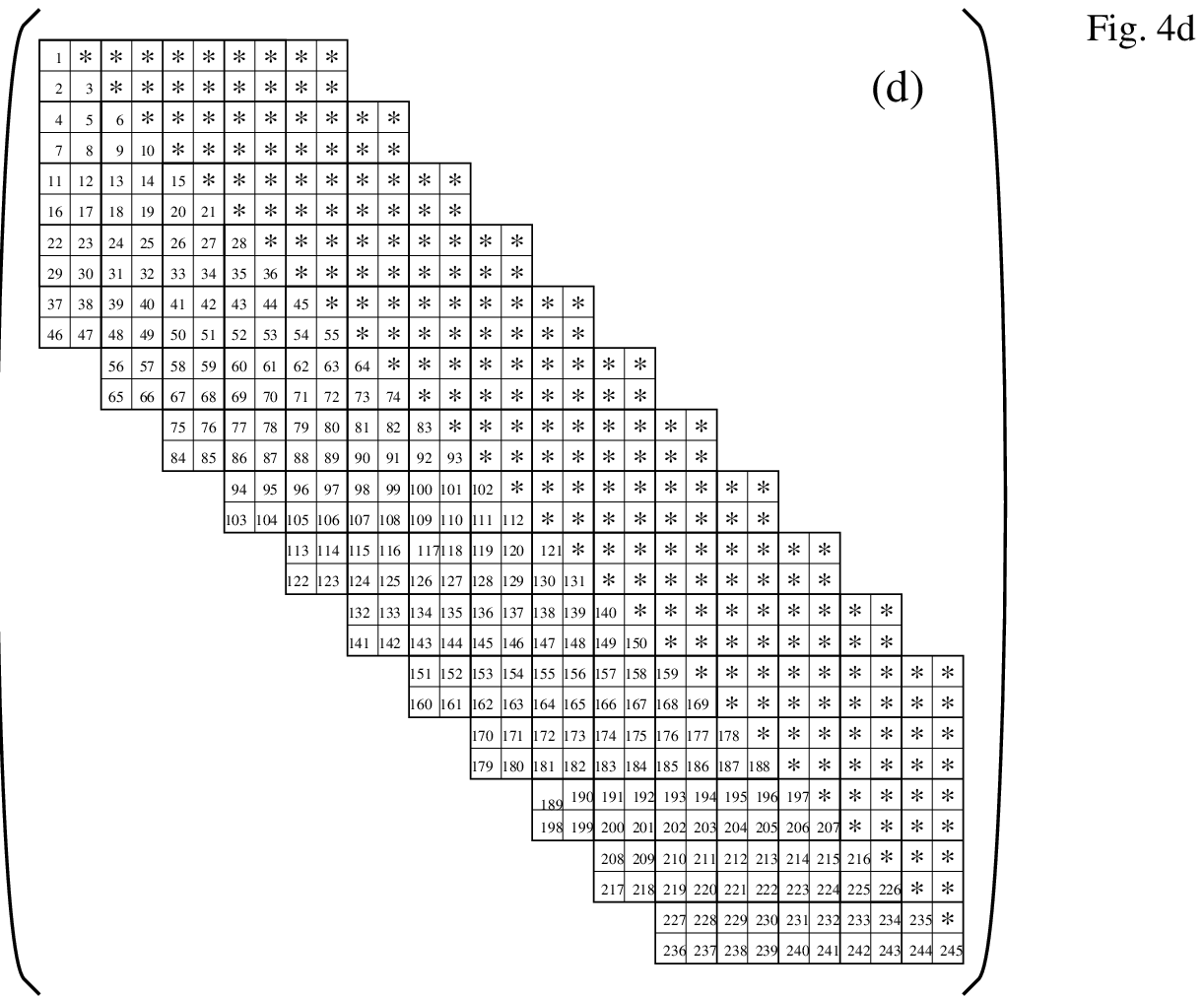} }
}
\end{figure}
\begin{figure}[H]
\centerline{
{ \epsfysize=6cm \epsfxsize=7cm                  
\epsffile{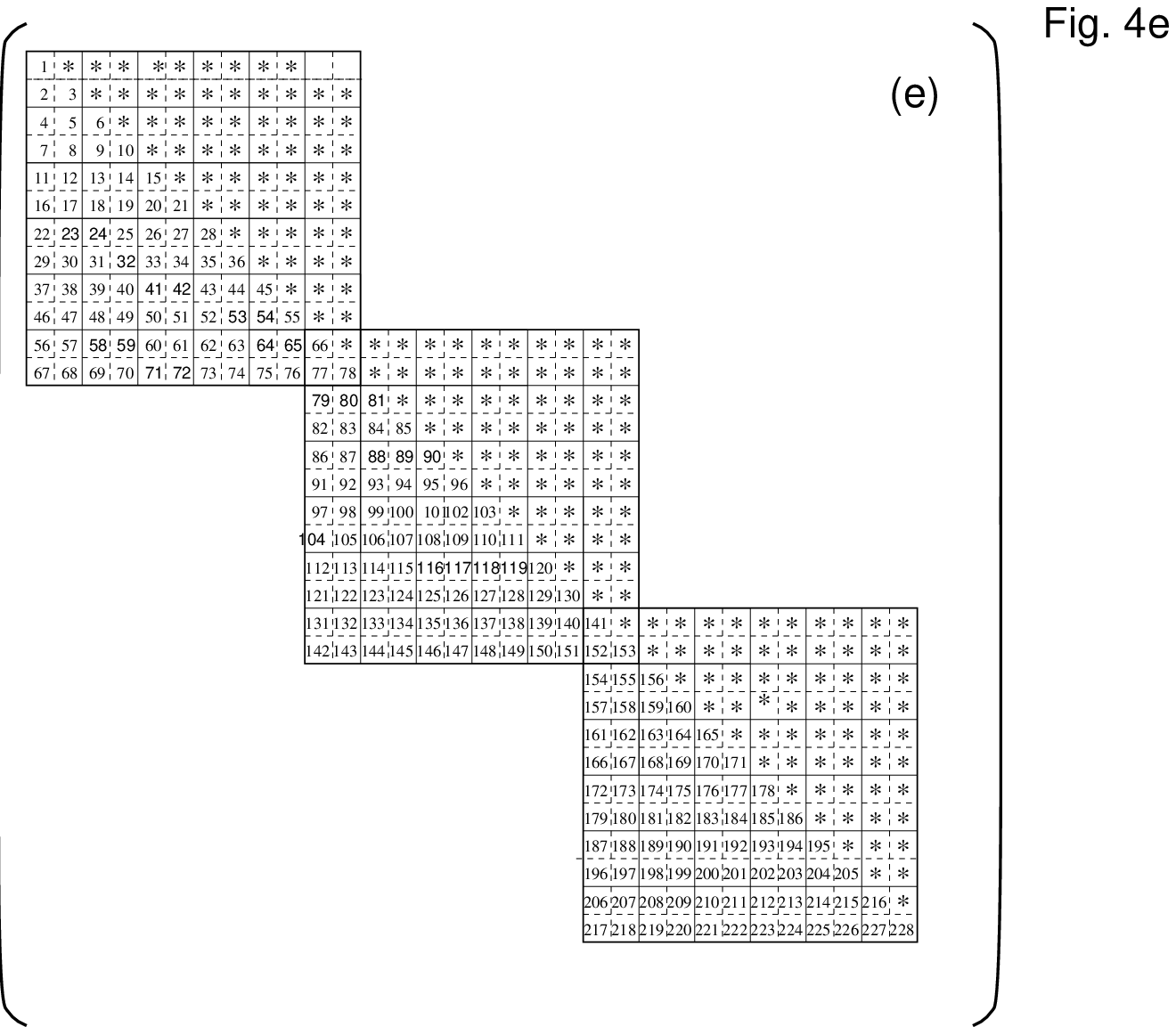} }          \hspace{2cm}
{ \epsfysize=6cm \epsfxsize=7cm
\epsffile{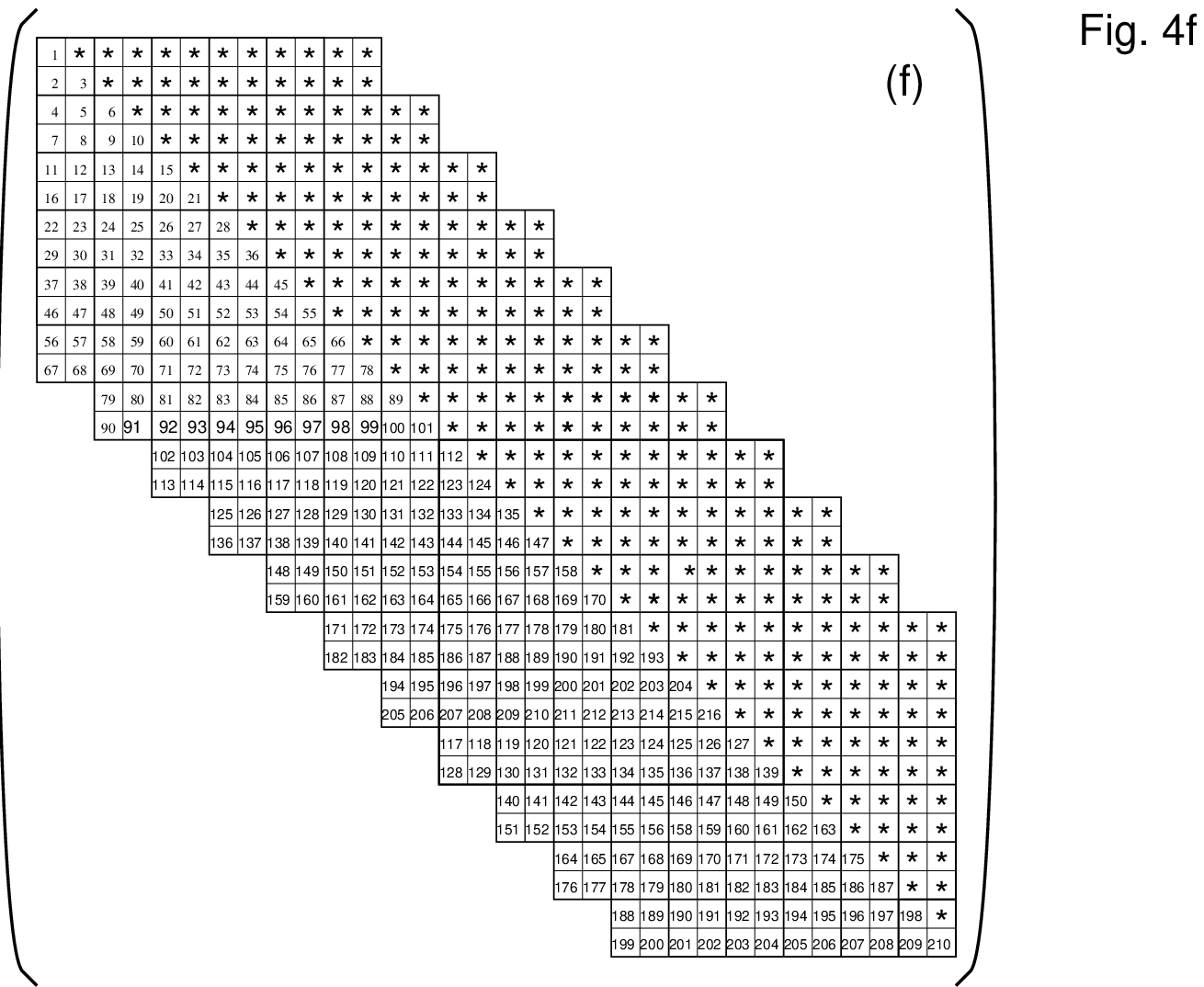} }
}
\vskip 0.5cm    
{\small {\bf Fig. 4:} \quad
Occupation patterns of stiffness matrices which result from the
finite element discretization of Eq. (\ref{Equ.2.1}). The figures
(a), (c), (e) display matrices which correspond to $3^{\rm rd}$ order, 
$4^{\rm th}$ order
and $5^{\rm th}$ order Lagrange type 
FEM. Figues (b), (d), (f) display patterns
which result from $3^{\rm rd}$ order, $4^{\rm th}$ order 
and $5^{\rm th}$ order B-spline FEM discretizations.
The numbers represent counter indices used in a vector storage technique.
}
\end{figure}
\vskip 0.5cm 
The boundary conditions at $r=0\,{\rm fm}$ depend on the quantum number
$\kappa $ and are defined in the following way.
\begin{eqnarray}
f(r=0)=0 \quad\mbox{\rm for}\quad \kappa = -1 \, \\
g(r=0)=0 \quad\mbox{\rm for}\quad \kappa = +1 \, \\
g(r=0)=0 \quad\mbox{\rm and}\quad f(r=0)=0 
\quad\mbox{\rm for}\quad \vert\kappa\vert > 1.
\label{Equ.4.8}
\end{eqnarray}
The system of algebraic equations (\ref{Equ.4.4})
forms a generalized eigenvalue problem of
the form $A\,u=\varepsilon\,N\,u$ with stiffness matrices $A$ and $N$
can be analyzed from the resulting matrix equation
\begin{eqnarray}
\Bigl[S_1\otimes\sigma_3\sigma_1+S_2\otimes\sigma_3\sigma_1-
      \kappa S_2\otimes\sigma_1+m S_3\otimes\sigma_3+
      S_4\otimes\sigma_3+
      S_5\otimes {\bf 1}_2 \Bigr]  u 
     = \varepsilon\Bigl[ S_3\otimes {\bf 1}_2 \Bigr]  u
\label{Equ.4.9}
\end{eqnarray}
where $u$ is a vector with components 
$(u_1^{(g)},u_1^{(f)},...,u_n^{(g)},u_n^{(f)} )^T$.
The symbols $S_1$ to $S_5$ denote the stiffness matrices of the operators
on the l.h.s. of Eq. (\ref{Equ.4.2}),
\begin{eqnarray}
S_1 &=&  \Bigl< w_{p'}(r)\Big\vert \partial_r \Big\vert B_p(r)\Bigr>, 
\nonumber\\
S_2 &=& \Bigl< w_{p'}(r)\Big\vert r^{-1} \Big\vert B_p(r) \Bigr>,
\nonumber\\
S_3 &=& \Bigl< w_{p'}(r) \Big\vert B_p(r)\Bigr>, 
\nonumber\\
S_4 &=& \Bigl< w_{p'}(r) \Big\vert S(r) \Big\vert B_p(r)\Bigr>, 
\nonumber\\
S_5 &=& \Bigl< w_{p'}(r) \Big\vert V(r)\Big\vert B_p(r)\Bigr>. 
\label{Equ.4.10}
\end{eqnarray}
In Figs. 4a-f, occupation patterns of the stiffness matrices A are
displayed for Lagrangian and B-spline finite element discretizations.
The matrices in Figs. 4a, 4c, 4e result from the Lagrange FEM with
$3^{\rm rd}$ order, $4^{\rm th}$ order and $5^{\rm th}$ order finite elements.
For comparison, I show the corresponding stiffness matrices of the
B-spline FEM in Figs. 4b, 4d, 4f. The sub-block structure of
$2\times 2$-blocks in all matrices results from the fact that
Eq. (\ref{Equ.2.1}) is a system of two coupled equations. 
The number of occupied $2\times 2$-blocks for a given order
$n^{\rm ord}$ in the Lagrange FEM is
$n^{fe}\cdot\bigl[ (n^{\rm ord})^2 + 2n\bigr]+1$ while
in the B-spline method the occupation increases to
$n^{fe}\cdot\bigl[ 2(n^{\rm ord})^2 + n\bigr]+1$. $n^{fe}$ denotes here
for both cases the number of finite elements used in the Lagrange method 
and is different from the number of elements which is used in the B-spline FEM.

The FEM discretization of the Klein-Gordon equations (\ref{Equ.2.4.a}) 
to (\ref{Equ.2.4.d}) is described in the appendix.
Finally, the coupled system of differential equations
(\ref{Equ.2.1}), (\ref{Equ.2.4.a}) to (\ref{Equ.2.4.d})
is replaced by a system of linear algebraic equations
\begin{equation}
A({\vec\sigma},{\vec\omega^{\,0}},{\vec\rho^{\,0}},{\vec A^{\,0}})\, u
 = \varepsilon\, N\, u
\label{Eq.4.12.a}
\end{equation}
for the node variables $u_p^{(g)},u_p^{(f)}$ of nucleon spinors, and
\begin{eqnarray}
B_{\sigma}({\vec\sigma}_{\rm old})\, {\vec\sigma}={\vec r}^{\rm (s)} \, \\
\label{Eq.4.12.b}
B_{\omega}\, {\vec\omega}^{\,0}={\vec r}^{\rm (v)}   \, \\
\label{Eq.4.12.c}
B_{\rho}\, {\vec\rho}^{\,0}={\vec r}^{\rm (3)}       \, \\
\label{Eq.4.12.d}
B_{A}\, {\vec A}^{\,0}={\vec r}^{\rm (em)}          
\label{Eq.4.12.e}
\end{eqnarray}
for the node variables $\sigma_p$, $\omega_p$, $\rho_p$, and $A_p$
of the meson fields
$\sigma(r)$, $\omega^0(r)$, $\rho^0(r)$, and photon field $A^0(r)$. 
The occupation patterns of the matrices $B_{\sigma},$ $B_{\omega},$
$B_{\rho},$ and $B_{A^0}$ for various shape functions
are very similar to those of the matrix A (Figs. 4a-d).
The main difference is that $2\times 2$-blocks have to be replaced
by single matrix elements.\hfill\break
%
%
%
%
%
\section {Analysis of spuriousity}
%
The appearance of spurious solutions in applications of 
FEM is a well known problem in general. First applications of the
finite element method in relativistic nuclear physics \cite{PVR.96}
have shown that spurious solutions appear in the spectra of the
Dirac equation. Linear finite elements have been used to calculate
solutions of the relativistic nuclear slab model. Comparisons of the
solutions with solutions that have been obtained with other numerical
techniques (shooting method) have shown that FEM reproduces the 
physical spectra very well and that spurious solutions have no
influence. In a further step \cite{PVRR.97} Lagrangian finite elements
of 1. to $4^{\rm th}$ order have been used in the self-consistent solution of the
RMF equations of sperical nuclei. In these calculations, it has been
shown (up to $4^{\rm th}$ order) that the density of spurious solutions in the
spectra decreases for increasing order of the elements. 

In this sections, I present a systematic study of the spurious spectra
which appear in the spherical symmetric case. In the initial step of
a self-consistent ground state calculation of $^{208}_{82}Pb$, Woods-Saxon
potentials
\begin{eqnarray}
\label{Equ.5.1.a}
S(r) = S(0) \Bigl(1+\exp({{r-r_s}\over a}))\Bigr)^{-1},  \\
\label{Equ.5.1.b}
V(r) = V(0) \Bigl(1+\exp({{r-r_s}\over a})\Bigr)^{-1},
\end{eqnarray}
are used for the scalar potential $S(r)$ and for the vector potential $V(r)$.
For $^{208}Pb$ the values of these potentials at $r=0\,{\rm fm}$ are chosen
$S(0)=-395\,{\rm MeV}$ and $V(0) = 320\,{\rm MeV}$, respectively. 
$a=0.5\,{\rm fm}$ and $r_s = 9.0\,{\rm fm}$. The calculation is performed
on a uniform radial mesh extending from $r_{\rm min} = 0\,{\rm fm}$
to $r_{\rm max} = 20\,{\rm fm}$. A smaller value $r_{\rm max} = 12\,{\rm fm}$
would be sufficient for $^{208}Pb$ to obtain good approximations of the
bound single particle states. For a good resolution of the continuum,
however, a large extension of the mesh in coordinate space is necessary.
An extremely high number of 200 mesh points has been used in the
calculation of the sprectra shown in Fig. 5a to Fig. 5f. The reason
for that choice will become clear from the subsequent discussion of Fig. 6. 
For a nucleon mass of $939\,{\rm MeV}$ (parameter set NL3), the Dirac gap
extends from $-939\,{\rm MeV}$ to $+939\,{\rm MeV}$. Bound solutions
are expected to have energies which are located in the Dirac gap. In the
following calculations an energy window ranging from $-1300\,{\rm MeV}$ to
$+1300\,{\rm MeV}$ has been chosen which covers parts of the lower and
upper continuum as well.

The results which are presented in the subsequent discussion correspond to the
first iteration step and a value $\kappa = -1$ (s-waves).
Spurious spectra of similar eigenvalue distributions are obtained for all other
$\kappa$-values ($\kappa = +1,\pm 2,\pm 3,...$). Disregarding the fact that
the eigensolutions change while they converge, very similar results are found
in all iteration steps of the selfconsistent iteration.

One of the most interesting questions to be answered in the present paper is,
whether the appearance of spurious solutions can be avoided using B-spline
finite elements instead of Lagrangian elements. Since both methods are
identical in the case of $1^{\rm st}$ order, spurious solutions 
appear also in the B-spline FEM. 
However, from that one can not conclude that spurious 
solutions appear in FEM discretizations with B-splines of higher order.
The following six figures Fig. 5a to Fig. 5f display energy spectra
of Eq. (\ref{Equ.2.1}) which have been calculated for many different orders
with both methods, the Lagrange FEM and the B-spline FEM.
In Fig. 5a and Fig. 5b the positive and negative spectra are shown
for $1^{\rm st}$ order to $4^{\rm th}$ order finite elements. The white circles correspond to
physical eigenvalues which have been calculated with Lagrange type elements.
They are located at the same energies as the white triangles which
correspond to physical eigenvalues obtained with the B-spline FEM.
The figures show that the physical spectra are independent on
the order and on the used method. 
However, the number of black filled circles and
filled triangles in the spectra decreases for increasing order of the
used shape functions. The filled symbols indicate eigenvalues of spurious
solutions. It turns out that the distributions of spurious eigenvalues
over the entire energy range are identical for both methods and in all
orders. In comparison to the Lagrange FEM, 
the B-spline method does obviously not reduce the number of spurious
states as long as the order is the same used in both methods. 
However, for both methods, the
density of spurious solutions in the spectra can be strongly reduced by
increasing the order of the finite elements. Particularly, from Fig. 5c to
Fig. 5f, on can see, that the spurious eigenvalues drift away from the
Dirac gap when the order is increased. Consequently, for any energy window
there exists an order which is high enough so that no spurious solutions
appear in the window.
An exception forms the region between the lowest positive physical eigenvalue
and the highest negative physical eigenvalue. For all element orders with
both methods, no spurious solution has been found in that region.

%
%
%
%
%
\begin{figure}[H]
\centerline{
{ \epsfysize=6cm \epsfxsize=7cm                  
\epsffile{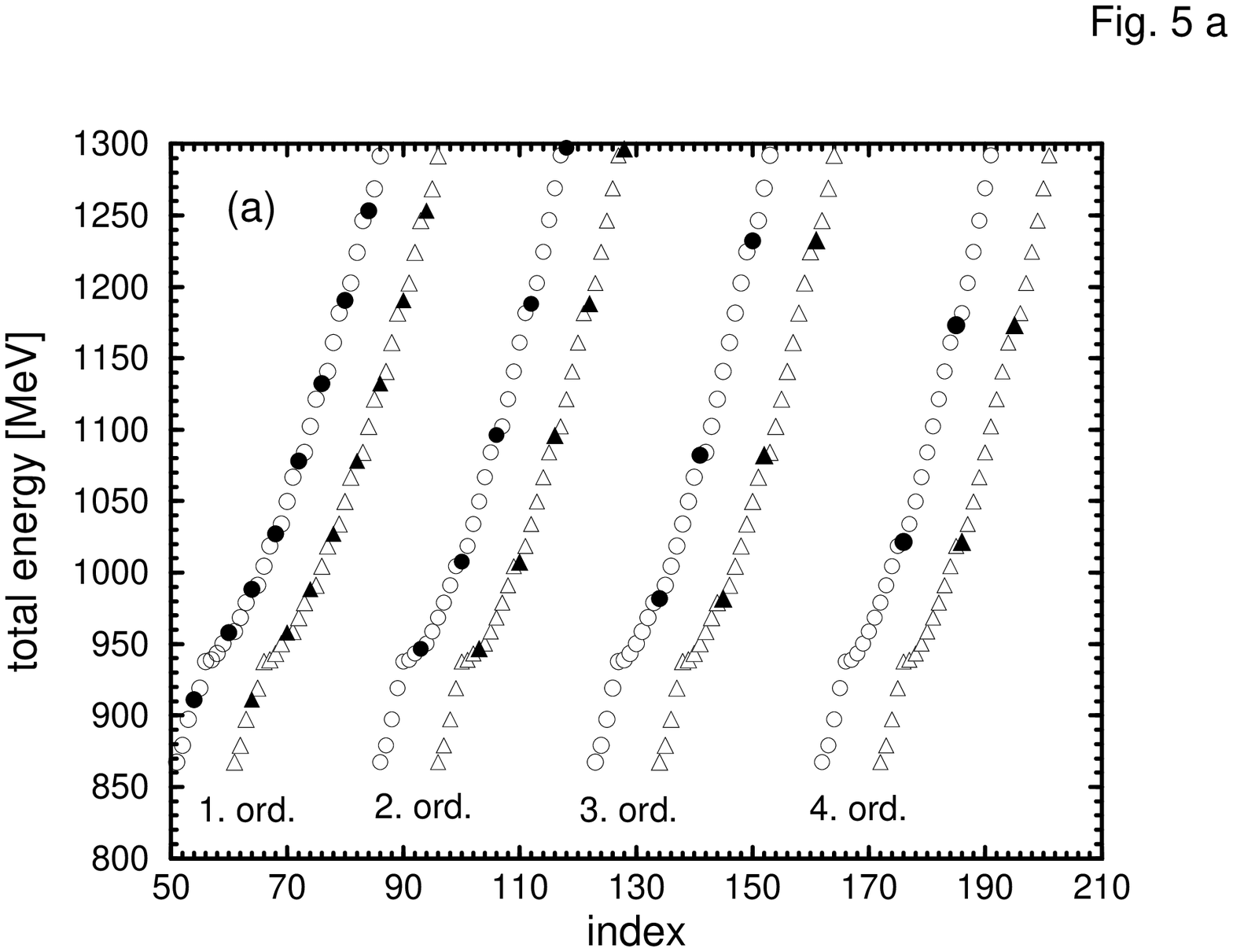} }          \hspace{1cm}
{ \epsfysize=6cm \epsfxsize=7cm                  
\epsffile{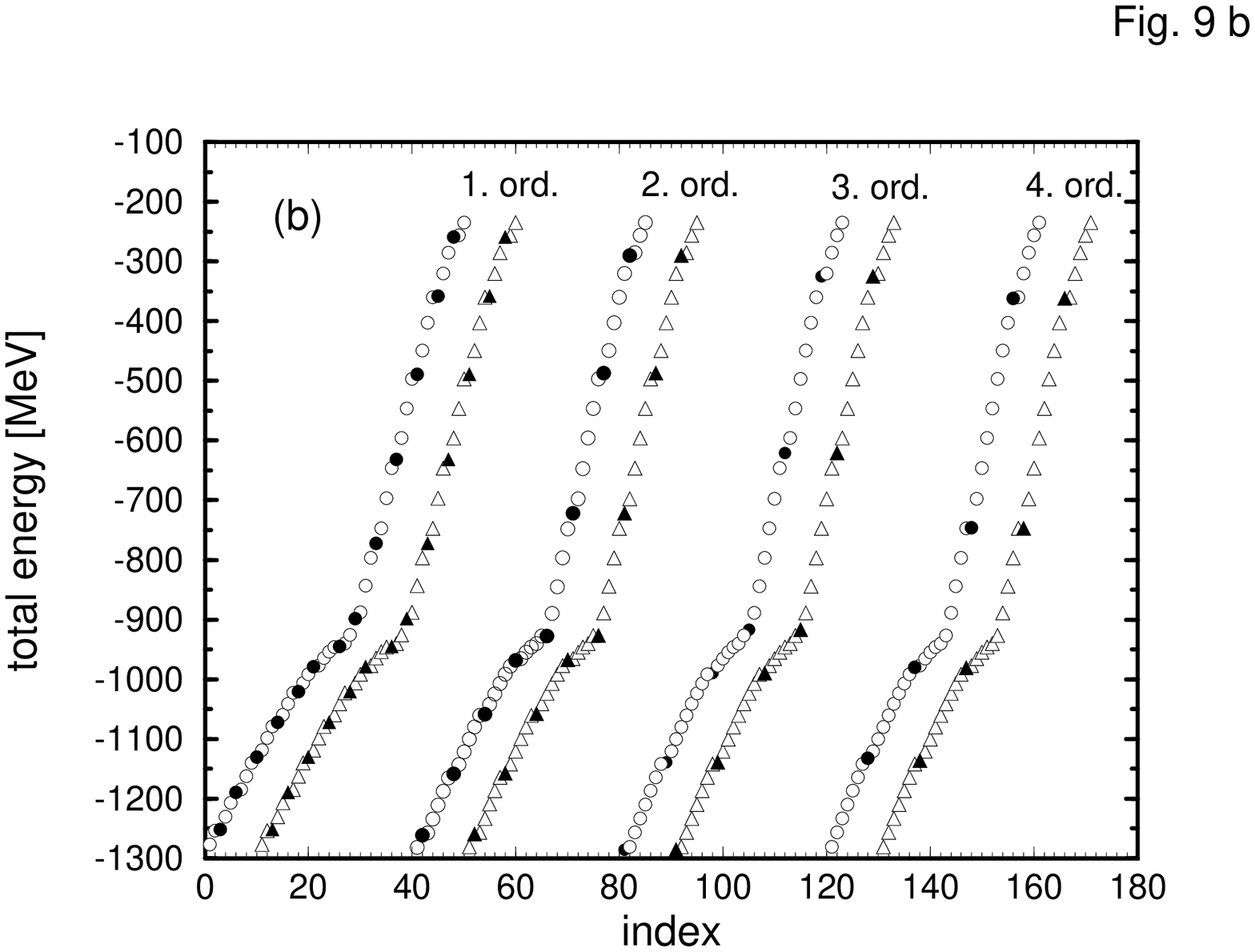} }
}
\end{figure}
\begin{figure}[H]
\centerline{
{ \epsfysize=6cm \epsfxsize=7cm                  
\epsffile{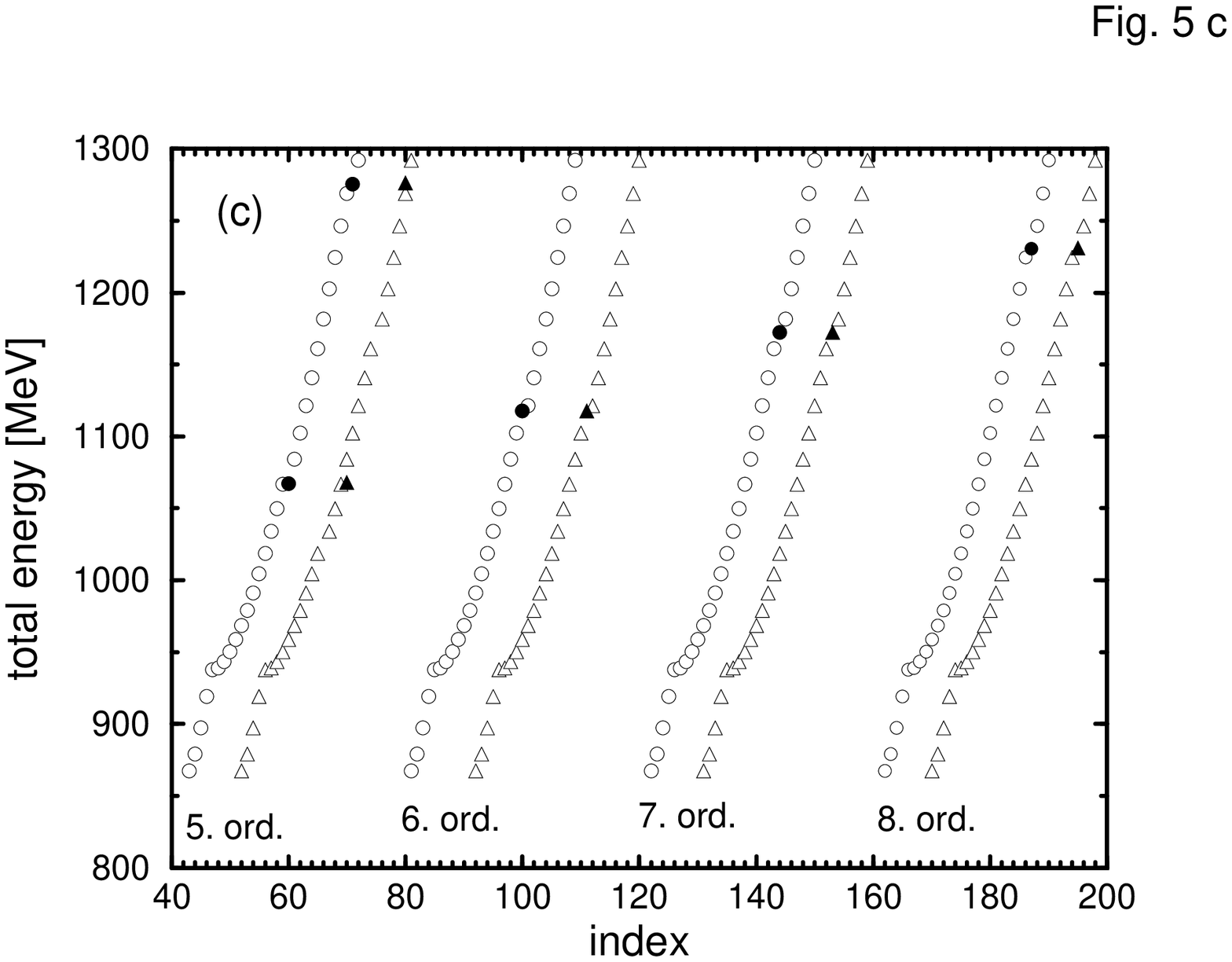} }          \hspace{1cm}
{ \epsfysize=6cm \epsfxsize=7cm                  
\epsffile{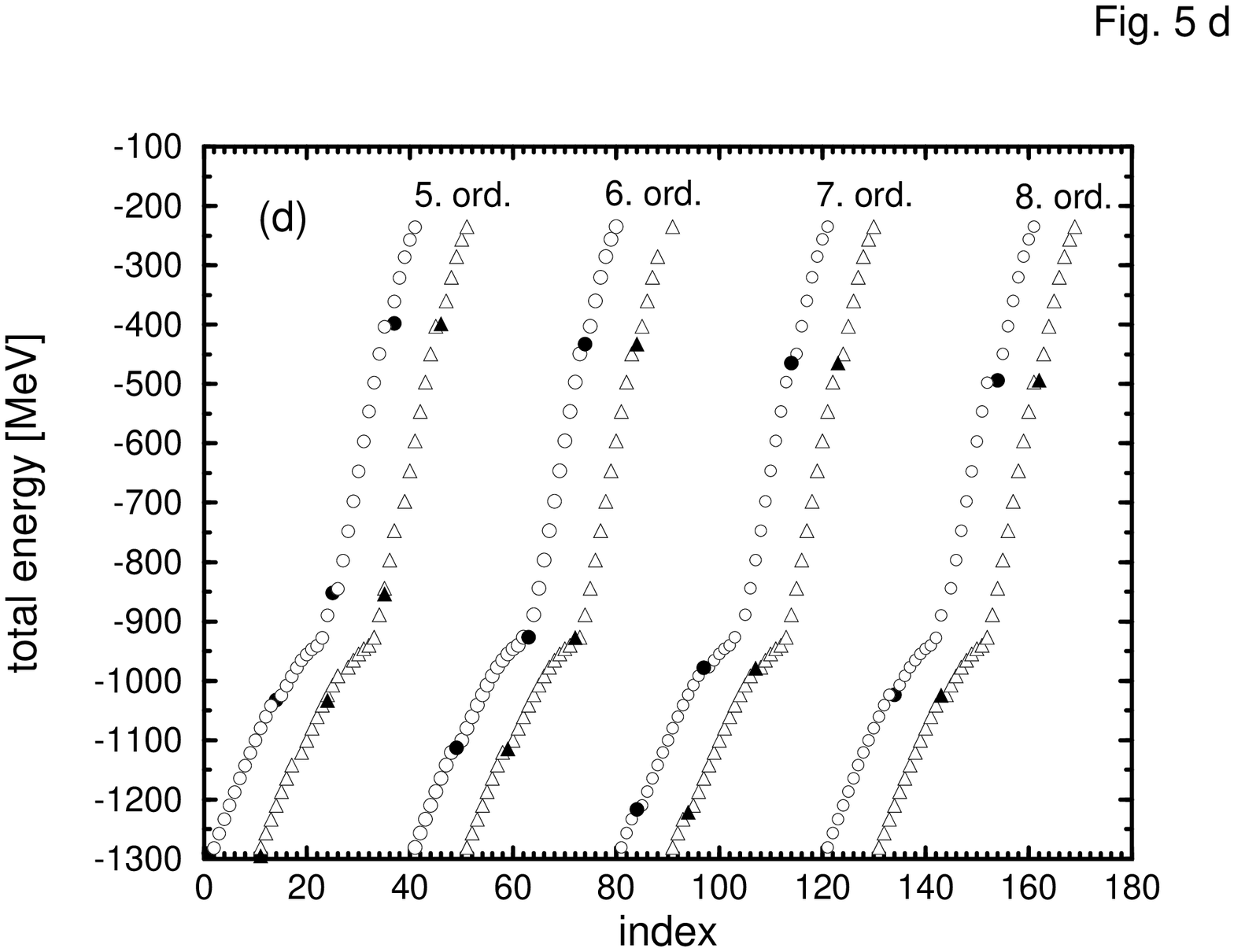} }
}
\end{figure}
\begin{figure}[H]
\centerline{
{ \epsfysize=6cm \epsfxsize=7cm                  
\epsffile{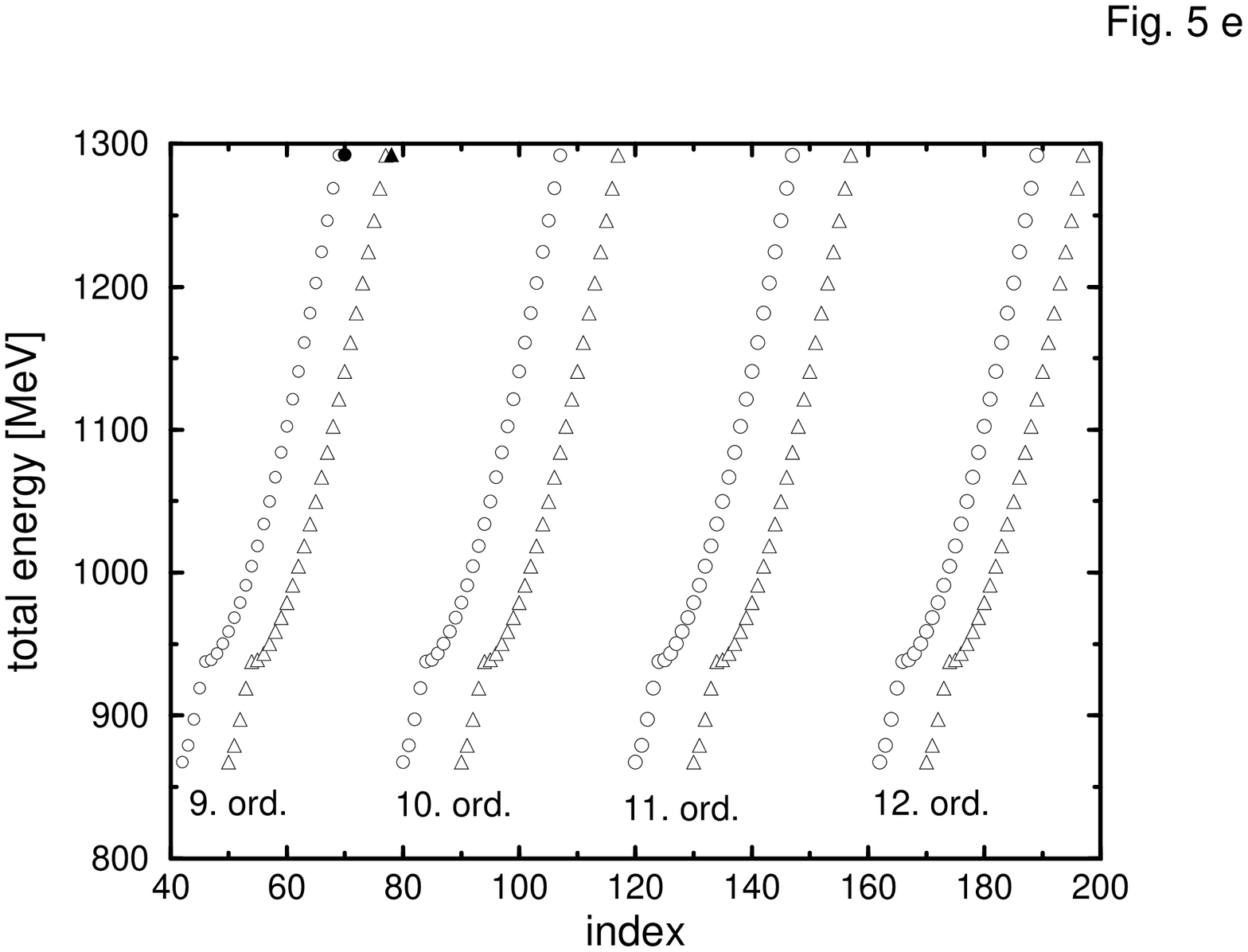} }          \hspace{1cm}
{ \epsfysize=6cm \epsfxsize=7cm
\epsffile{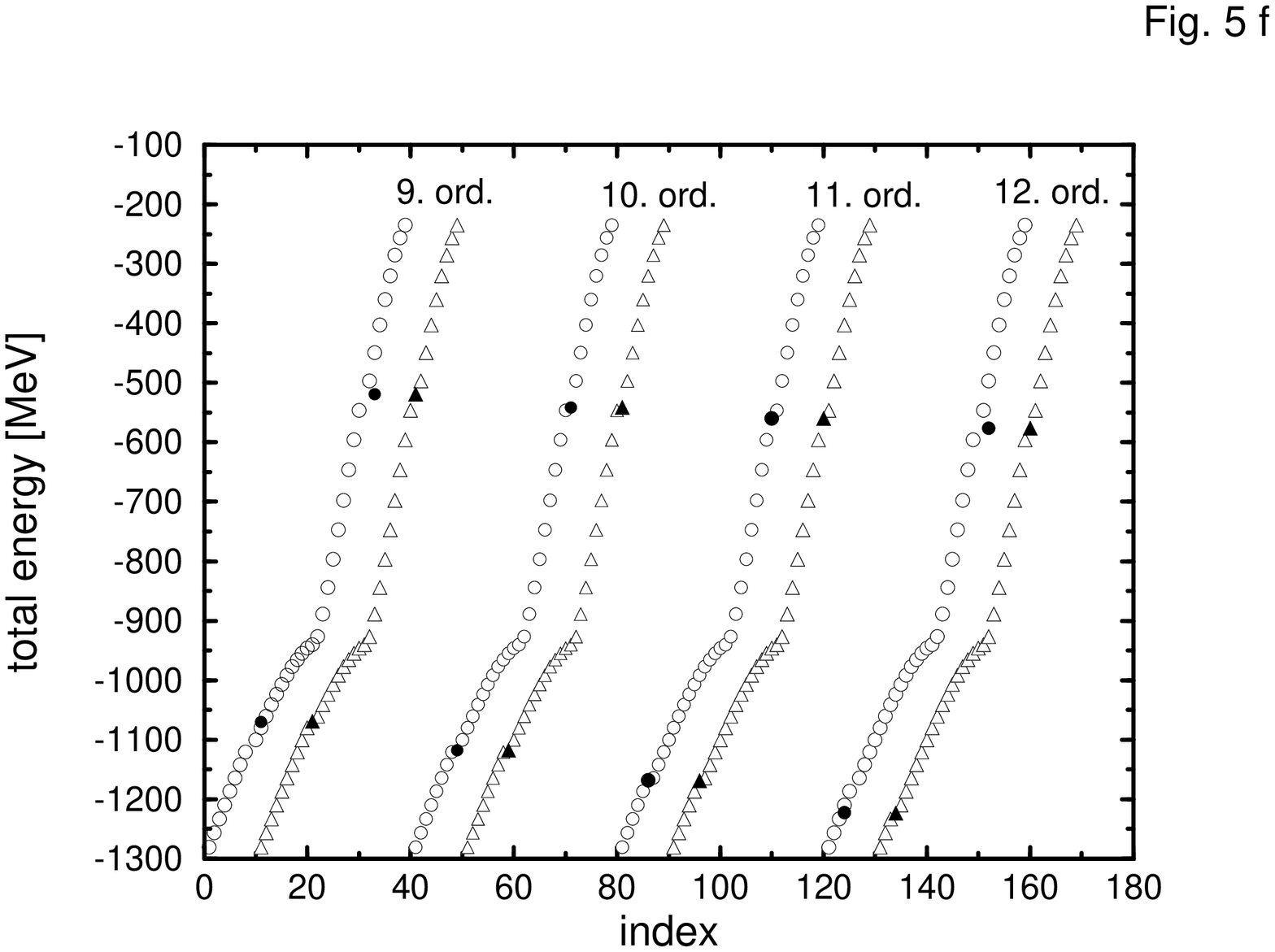} }
}
\vskip 0.5cm    
{\small {\bf Fig. 5:} \quad
Energy eigenvalues of the Dirac equation (\ref{Equ.2.1}) for the case $\kappa = -1$.
The spectra are compared for the Lagrange FEM (circles) and the B-spline FEM
(triangles). The used finite element orders are indicated in the figures.
Filled symbols correspond to spurious eigenvalues. All eigenvalues which appear
in the energy window $\bigl[-1300{\rm MeV},\,+1300{\rm MeV}\bigr]$ 
are displayed for $1^{\rm st}$ order to $12^{\rm th}$ order elements. 
}
\end{figure}
It is also important to analyze the dependence of the spurious spectrum
on the number of mesh points. In Fig. 6, the number of spurious solutions
is displayed for different orders as a function of the number of mesh points
in a constant radial box 
($r_{\rm min} = 0 \,{\rm fm},\, r_{\rm max} = 20\, {\rm fm}$). The results
show that the number of spurious solutions is independent on the number of
mesh points if this number is sufficiently large. This is true for all orders
of finite elements. The solid lines in Fig. 6 show the results which have
been obtained for the above defined Woods-Saxon potentials. For all
finite element orders the number of spurious states in the above defined
energy window increases at low mesh point numbers and decreases
monotonically at high mesh point numbers. At large numbers of mesh points
(''asymptotic region'') the number of spurious solutions is constant for all
element orders. This has been tested up to the very large number of 
600 mesh points but is not shown in the figure. 
For the calculation of the spectra shown in Fig. 5a to 5b,
I have used a number of mesh points (200) which is in that asymptotic region
to make sure that they (in particular the spurious spectra) are independent
on the number of mesh points.

An explanation for the curves in Fig. 6 is found with the concept of
Sobolev space. In reference \cite{PVRR.97} it has been outlined that
a Sobolev space $W_p^m(\Omega)$ is a completion of the test function
space $C_0^{\infty}(\Omega)$ with respect to the Sobolev norm
$\Vert\cdot\Vert_{m,p}$ defined in Eq. (\ref{Equ.4.1}). Thus,
$C_0^{\infty}(\Omega)\subset W_p^m(\Omega)$ for all integer numbers
$m\ge 0$. All spaces $W_p^m(\Omega)$, where $m > 0$, are subspace of the
largest Sobolev space $W_p^0(\Omega)$ and
\begin{equation}
W_p^{m+1}(\Omega)\subset W_p^{m}(\Omega)\qquad\mbox{for all}\qquad m\ge 0.
\label{Equ.5.3}
\end{equation}
The shape functions of $m^{\rm th}$ order finite elements are element of
$W_p^m(\Omega)$ but the shape functions of any lower order finite elements
are not in $W_p^m(\Omega)$. In finite element discretizations of low order
$m$ is small and one works in a correspondingly large space $W_p^m(\Omega)$.
The weak form, of a differential equation, expressed in terms of the
weighted residual, allows more solutions than the solutions of the
original problem. All solutions which are found for a certain FEM order $m$
are element of $W_p^m(\Omega)$. For increasing order $m$ the space 
$W_p^m(\Omega)$ shrinks and the number of spurious solutions in the
weak form is reduced while all physical solutions are maintained.
This is seen from Fig. 5a to Fig. 5f for one large number of mesh points.
It explains in general the reduction of the number of unphysical solutions
in Fig. 6 for all mesh point numbers when the order of the FEM-ansatz is
increased. For a uniform finite element mesh with constant width $h$, the
$m^{\rm th}$ order shape functions of the whole FEM discretization span up
a space $S_h^{m}(\Omega)$. Starting from an initial discretization where
$h_0$ is large, a sequence of spaces $S_{h_i}^m(\Omega)\, i=0,1,2,...$
is generated when the mesh is refined for increasing index $i$, where
$h_{i+1}< h_i$. The direct sum of the spaces $S_{h_i}^m(\Omega)$ 
converges against $W_p^m(\Omega)$ and thus 
$\bigoplus\limits_{i=0}^{\infty}S_{h_i}^m(\Omega)=W_p^m(\Omega)$.
Different spaces $S_{h_i}^m(\Omega)$ and $S_{h_j}^m(\Omega)$ where $i\ne j$ 
can have non-trivial intersection. There are even cases where
$S_{h_i}^m(\Omega)\subset S_{h_j}^m(\Omega)$ when $h_i < h_j$. In the
example of the two spaces $S_h^1(\Omega)$ and $S_{h/2}^1(\Omega)$ it is
obvious that $S_h^1(\Omega)\subset S_{h/2}^1(\Omega)$ since each
linear shape function which is basis function in $S_h^1(\Omega)$ can
be represented as linear combination of shape functions (basis functions)
of $S_{h/2}^1(\Omega)$.
The strong increase of the graphs in Fig. 6 at small mesh point numbers,
where $h$ is large, is explained by the fact that the spaces $S_h^m(\Omega)$
become large for decreasing $h$. The number of spurious solutions which
appear in $S_h^m(\Omega)$ increases simultaneously. However, there is a
second effect which is superposed to this first one.
%
%
%
%
%
\begin{figure}[H]
\centerline{
{ \epsfysize=12cm \epsfxsize=16cm                  
\epsffile{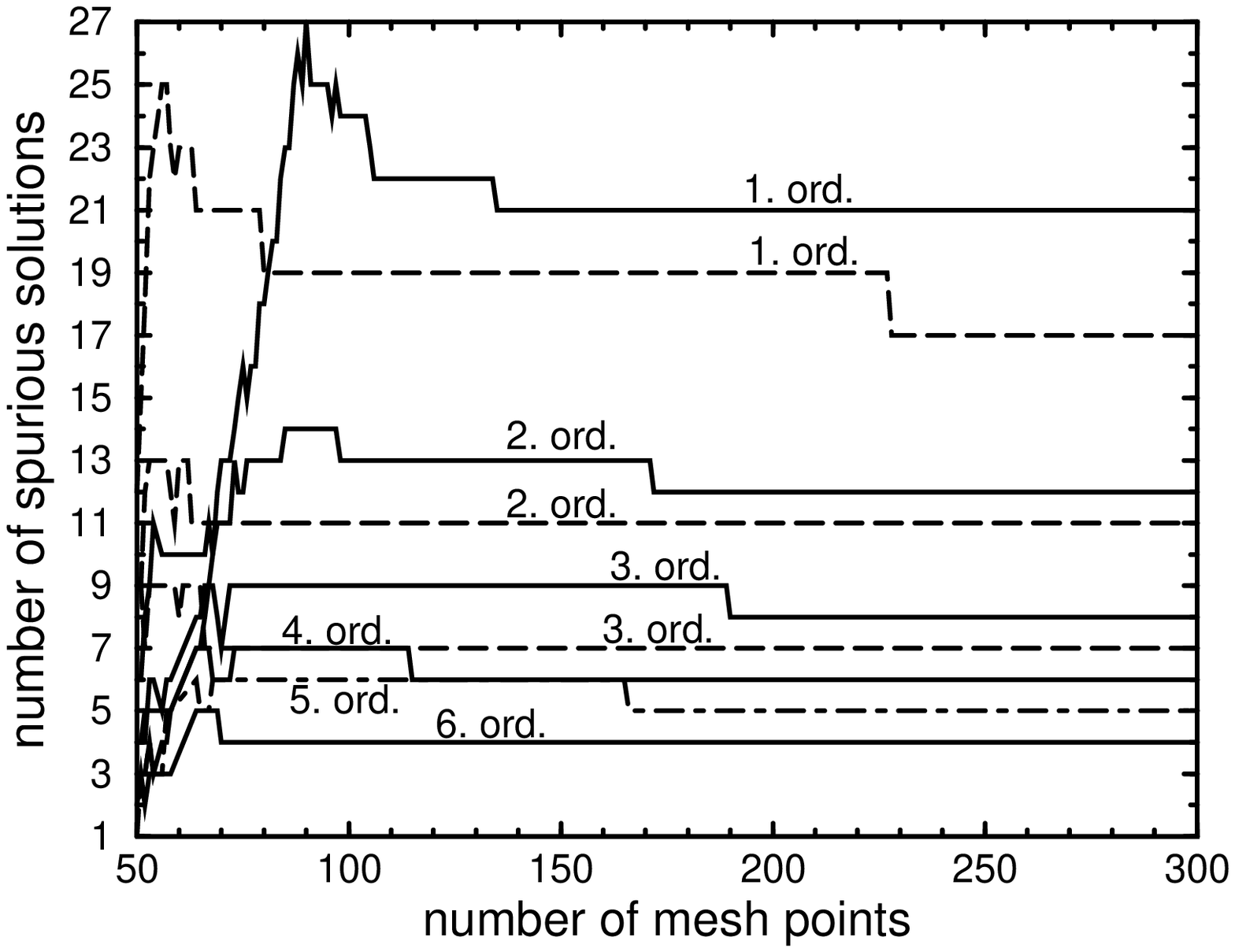} }
}
\vskip 0.5cm    
{\small {\bf Fig. 6:} \quad
Dependencies of the number of spurious solutions on the number of
mesh points are shown for $1^{\rm st}$ order to $6^{\rm th}$ order finite elements.
A constant mesh size ranging from $0\,{\rm fm}$ to $20\,{\rm fm}$ has
been used in the calculations. The solid (and dot-dashed) lines show 
results obtained for Woods-Saxon potentials $V(r)$ and $S(r)$. 
The dashed lines show corresponding results for zero potential.
}
\end{figure}
Spurious solutions which have a very high number of oscillations can only be
completely resolved in spaces $S_h^m(\Omega)$ where $h$ is small. However,
spurious states with high frequency can appear in subspaces 
$S_{\tilde h}^m(\Omega)$ where $\tilde h = \nu\cdot h\,(\nu = 1,2,3,...)$
and where only a fraction of the oscillations is resolved. Since the
corresponding kinetic energy which contributes to the total energy is small,
these solutions appear in the above chosen energy window. If the number of
mesh points is increased, additional oscillations are resolved and the
kinetic energy increases correspondingly. The corresponding total energy
appears no longer in the energy window. In the negative energy range such
solutions are shifted further into the negative continuum.

In Fig. 7a and Fig. 7b, spurious energy spectra are displayed for 
many different mesh point numbers. First order finite elements have been used.
The mesh point numbers have been chosen around the maximum of the solid curve 
in Fig. 6 which corresponds to first order.
The solid lines connect spurious eigenvalues which appear at constant
mesh point number which is indicated by the numbers atop of each line.  
%
%
%
%
%
\begin{figure}[H]
\centerline{
{ \epsfysize=7cm \epsfxsize=8cm                  
\epsffile{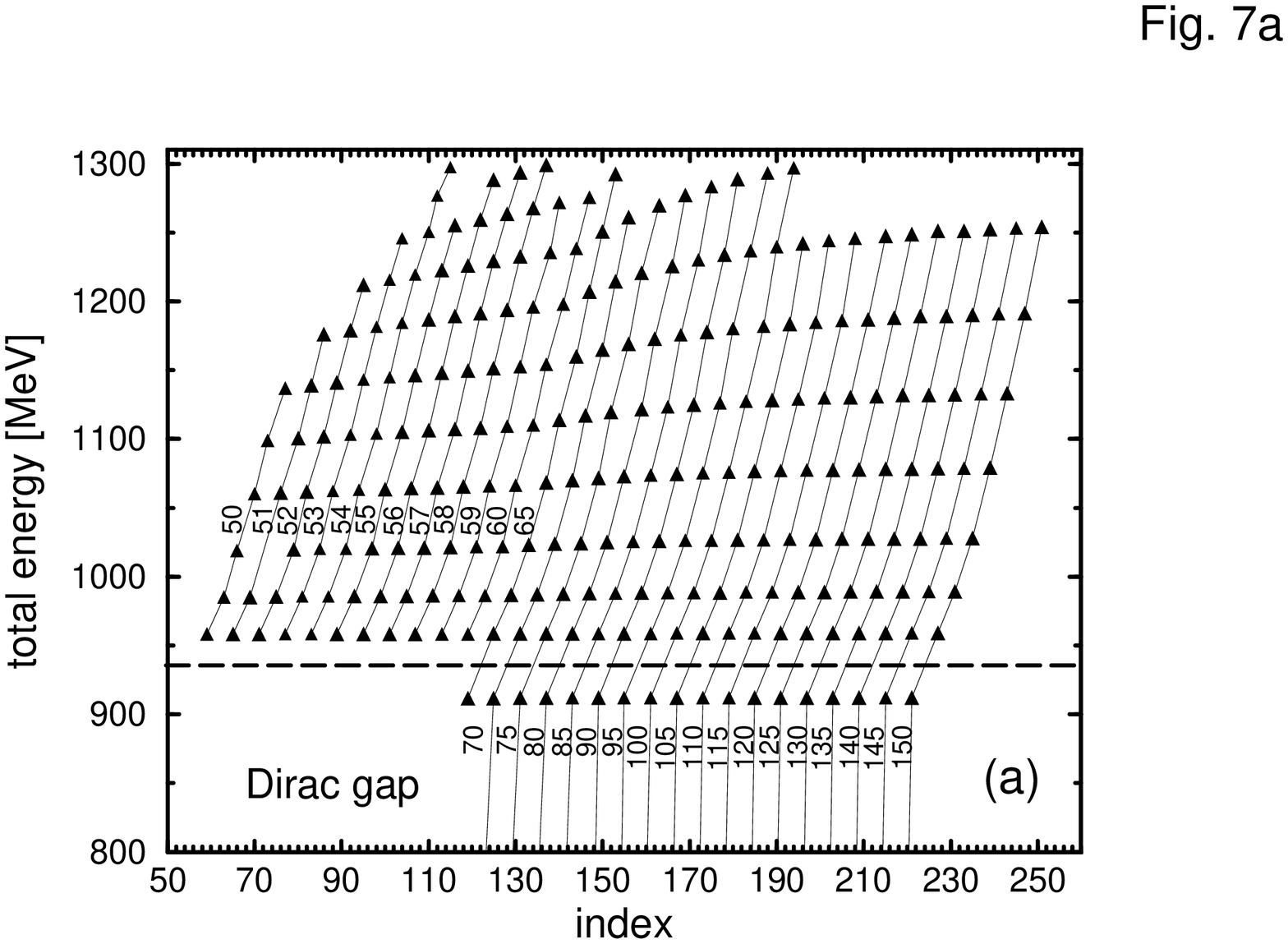} }          \hspace{1cm}
{ \epsfysize=7cm \epsfxsize=8cm                  
\epsffile{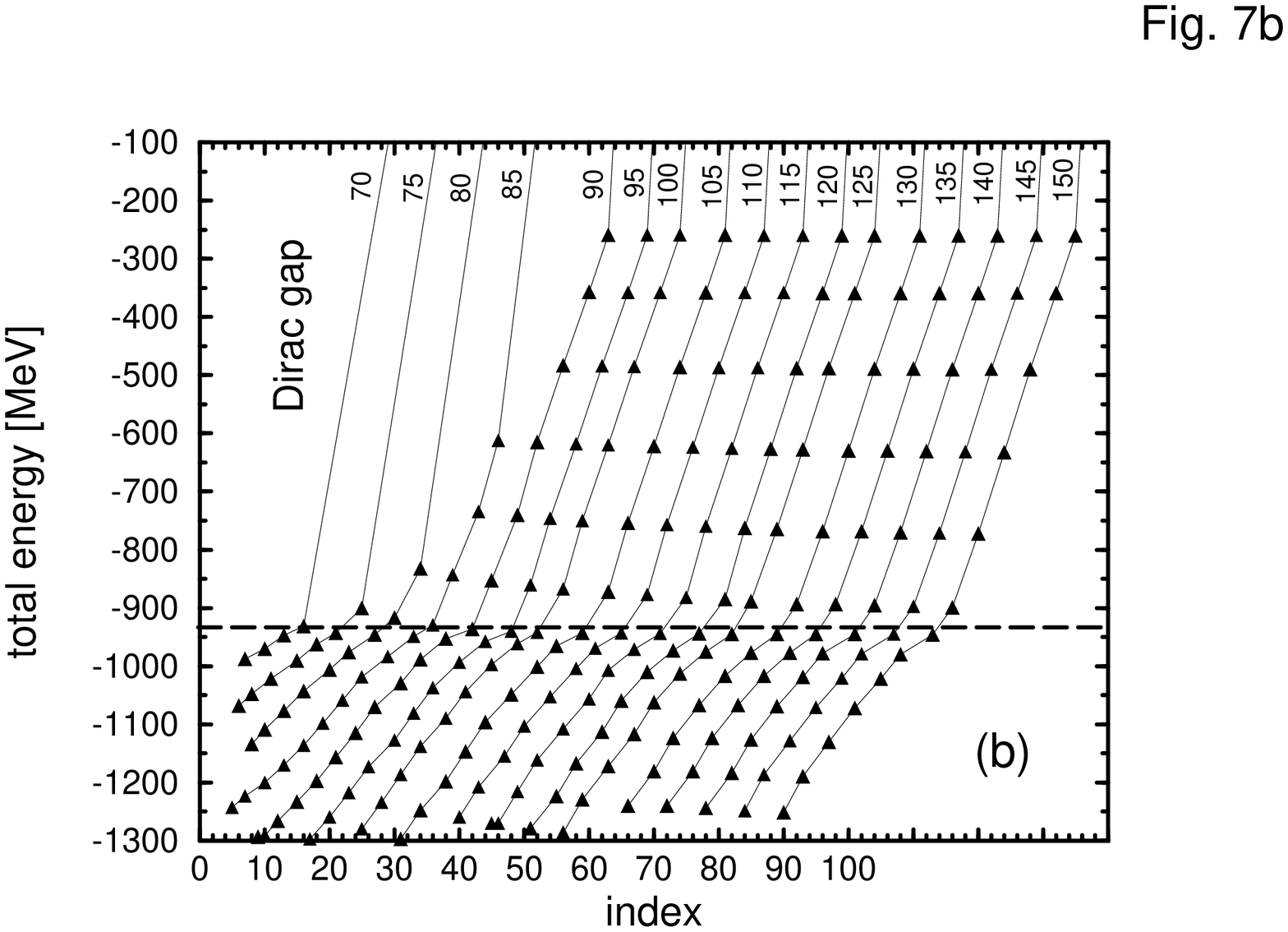} }
}
\vskip 0.5cm    
{\small {\bf Fig. 7:} \quad
Spurious spectra of the first order finite element discretization
are displayed for different mesh point numbers. The solid lines
connect eigenvalues which belong to the same discretization. The
number of mesh points is indicated above each line.
}
\end{figure}
%
%
%

%
%
%
\section {Numerical precision}
%
%
In the following, I present an analyse of the numerical precision of both
methods and compare the results. The quality in the approximation of the
exact solutions of (\ref{Equ.4.1}) depends essentially of the order of the
FEM-ansatz and on the number of mesh points used in a given domain
$\Omega = \bigl[ r_{\rm min},\,r_{\rm max}\bigr]$. In the subsequent
tables, neutron single particle eigenvalues are listed systematically
for increasing number of mesh points and increasing order of used finite
elements.
The eigenvalues correspond to solutions which have been obtained in the
initial step of a selfconsistent calculation for $^{40}{\rm Ca}$. In the first
iteration step, Woods-Saxon potentials of the form (\ref{Equ.5.1.a}) and
(\ref{Equ.5.1.b}) have been used with parameters $S(0)=-395\,{\rm MeV}$,
$V(0)=320\,{\rm MeV}$, $a=0.5\,{\rm fm}$ and $r_s=6.0\,{\rm fm}$. The mesh
size has been kept fixed with boundaries at $r_{\rm min} = 0\,{\rm fm}$
and $r_{\rm max}= 10\,{\rm fm}$.
In Table 1a neutron single particle energies which have been calculated
with linear finite elements are shown for the initial Woods-Saxon potential.
For increasing number of mesh points (see left column), the number of
unchanged decimal places reaches 8 at 200 mesh points. A comparison with
the last row of Table 1b shows that for linear elements the last digit
(decimal place 10) has not stabilized at the extremely large mesh point
number 600. Table 1b displays results that have been calculated with finite
elements of $3^{\rm rd}$ order. Between 109 and 121 mesh points (36-40 elements),
the results have stabilized in all 10 digits. 
In Table 1c, I show the corresponding results which have been obtained with
finite elements of $4^{\rm th}$ order. To demonstrate the enormous improvement in the
precision, the results are displayed up to 12 digits. Between 81 and 93
mesh points the $8^{\rm th}$ digit becomes stable and between 125 and
145 mesh points the precision achieves 12 digits.
%
%
%
%
%
%
%

{ \small
\begin{table}[H]
\begin{center}
\begin{tabular}{|l|l|l|l|l|l|l|}
\multicolumn{7}{l}{\bf 1. Order} \\
\hline\hline 
           &             &             &             &             &             & \\
 $n_{nod}$ & $E_{1s1/2}$ & $E_{1p3/2}$ & $E_{1p1/2}$ & $E_{1d5/2}$ & $E_{2s1/2}$ & 
 $E_{1d3/2}$  \\
\hline\hline
  10  &65.88844974 &57.33594611 &56.42536540 &47.45864481 &42.33620819 &45.34520938 \\ 
  20  &65.88133752 &57.31827220 &56.39907816 &47.42408909 &42.19281200 &45.28981141 \\
  30  &65.88108502 &57.31763456 &56.39825020 &47.42279518 &42.18737854 &45.28804244 \\
  40  &65.88104432 &57.31753150 &56.39811692 &47.42258515 &42.18648990 &45.28775645 \\
  50  &65.88103335 &57.31750369 &56.39808101 &47.42252842 &42.18624969 &45.28767927 \\
  60  &65.88102944 &57.31749377 &56.39806819 &47.42250815 &42.18616380 &45.28765171 \\
  70  &65.88102777 &57.31748952 &56.39806272 &47.42249948 &42.18612708 &45.28763993 \\
  80  &65.88102696 &57.31748747 &56.39806006 &47.42249529 &42.18610928 &45.28763422 \\
  90  &65.88102653 &57.31748638 &56.39805866 &47.42249306 &42.18609982 &45.28763119 \\
  100 &65.88102628 &57.31748575 &56.39805785 &47.42249178 &42.18609441 &45.28762946 \\
  150 &65.88102591 &57.31748480 &56.39805662 &47.42248983 &42.18608615 &45.28762681 \\
  200 &65.88102584 &57.31748464 &56.39805641 &47.42248951 &42.18608476 &45.28762637 \\
  300 &65.88102582 &57.31748458 &56.39805634 &47.42248938 &42.18608424 &45.28762620 \\
  600 &65.88102582 &57.31748457 &56.39805632 &47.42248936 &42.18608412 &45.28762616 \\
\hline \\
\end{tabular}
\end{center}
\vspace{0.4cm}
{\small {\bf Table 1a:} \quad
Neutron single particle energies in units of MeV 
which have been calculated with linear 
finite elements of Lagrange type. 
The number of used mesh points has been gradually increased as shown
in the left column. 
}
\end{table}
\vskip 0.4cm
\begin{table}[H]
\begin{center}
\begin{tabular}{|l|l|l|l|l|l|l|}
\multicolumn{7}{l}{\bf 3. Order} \\
\hline\hline 
           &             &             &             &             &             & \\
 $n_{nod}$ & $E_{1s1/2}$ & $E_{1p3/2}$ & $E_{1p1/2}$ & $E_{1d5/2}$ & $E_{2s1/2}$ & 
 $E_{1d3/2}$  \\
\hline\hline
  10 &65.87761838 &57.30204792 &56.38118314 &47.38090127 &42.16223936 &45.24653386  \\
  19 &65.88080395 &57.31691348 &56.39762680 &47.42131836 &42.18517027 &45.28703195  \\
  25 &65.88103067 &57.31748586 &56.39806263 &47.42248096 &42.18636302 &45.28761163  \\
  31 &65.88103176 &57.31749678 &56.39809005 &47.42250968 &42.18614523 &45.28768121  \\
  40 &65.88102636 &57.31748582 &56.39805841 &47.42249169 &42.18609035 &45.28762987  \\
  46 &65.88102599 &57.31748496 &56.39805720 &47.42249002 &42.18608570 &45.28762767  \\
  52 &65.88102587 &57.31748468 &56.39805656 &47.42248956 &42.18608460 &45.28762657  \\
  61 &65.88102583 &57.31748460 &56.39805637 &47.42248941 &42.18608424 &45.28762624  \\  
  70 &65.88102582 &57.31748458 &56.39805633 &47.42248937 &42.18608415 &45.28762619  \\
  79 &65.88102582 &57.31748457 &56.39805632 &47.42248936 &42.18608413 &45.28762617  \\
  91 &65.88102582 &57.31748457 &56.39805632 &47.42248936 &42.18608412 &45.28762616  \\
  100&65.88102582 &57.31748457 &56.39805632 &47.42248936 &42.18608412 &45.28762616  \\
  109&65.88102582 &57.31748457 &56.39805632 &47.42248935 &42.18608412 &45.28762616  \\
  121&65.88102581 &57.31748457 &56.39805632 &47.42248935 &42.18608412 &45.28762616  \\
\hline 
\end{tabular}
\end{center}
\vspace{0.4cm}
{\small {\bf Table 1b:} \quad
Same as Table 1a but for Lagrange FEM of $3^{\rm rd}$ order.
}
\end{table}
\vskip 0.4cm
\begin{table}[H]
\begin{center}
\begin{tabular}{|l|l|l|l|l|l|l|}
\multicolumn{7}{l}{\bf 4. Order} \\
\hline\hline 
           &             &             &             &             &             & \\
 $n_{nod}$ & $E_{1s1/2}$ & $E_{1p3/2}$ & $E_{1p1/2}$ & $E_{1d5/2}$ & $E_{2s1/2}$ & 
 $E_{1d3/2}$  \\
\hline\hline
   9  &65.8743670393&57.3047488634&56.3996551655&47.4029580598&42.0645085975&45.2829901985\\
  21  &65.8810348458&57.3174942083&56.3981447843&47.4224876123&42.1861879711&45.2877368920\\
  29  &65.8810254734&57.3174829474&56.3980580463&47.4224848157&42.1860824080&45.2876279606\\
  41 &65.8810257955&57.3174845038&56.3980562850&47.4224891983&42.1860839907&45.2876260664\\
  49 &65.8810258097&57.3174845532&56.3980563048&47.4224893238&42.1860840610&45.2876261362\\
  61 &65.8810258137&57.3174845640&56.3980563150&47.4224893479&42.1860841046&45.2876261556\\
  73 &65.8810258146&57.3174845663&56.3980563177&47.4224893526&42.1860841130&45.2876261605\\
  81 &65.8810258148&57.3174845667&56.3980563182&47.4224893535&42.1860841146&45.2876261614\\
  93 &65.8810258149&57.3174845669&56.3980563185&47.4224893539&42.1860841154&45.2876261619\\
 101 &65.8810258149&57.3174845670&56.3980563186&47.4224893540&42.1860841156&45.2876261620\\
 109 &65.8810258149&57.3174845670&56.3980563186&47.4224893541&42.1860841157&45.2876261621\\
 125 &65.8810258149&57.3174845670&56.3980563186&47.4224893541&42.1860841158&45.2876261621\\
 145 &65.8810258149&57.3174845670&56.3980563186&47.4224893541&42.1860841158&45.2876261622\\
\hline 
\end{tabular}
\end{center}
\vspace{0.4cm}
{\small {\bf Table 1c:} \quad
Same as Table 1b but for $4^{\rm th}$ order Lagrange FEM.
}
\end{table}
\vskip 0.4cm
\begin{table}[H]
\begin{center}
\begin{tabular}{|l|l|l|l|l|l|l|}
\multicolumn{7}{l}{\bf 5. Order} \\
\hline\hline 
           &             &             &             &             &             & \\
 $n_{nod}$ & $E_{1s1/2}$ & $E_{1p3/2}$ & $E_{1p1/2}$ & $E_{1d5/2}$ & $E_{2s1/2}$ & 
 $E_{1d3/2}$  \\
\hline\hline
  11 &65.8784571165&57.3086583184&56.3940420670&47.4017253841&42.1771582984&45.2818654596 \\
  21 &65.8810087294&57.3174496620&56.3980829394&47.4224368621&42.1858924489&45.2876670981 \\
  31 &65.8810255741&57.3174839393&56.3980557441&47.4224881501&42.1860830079&45.2876254614 \\
  41 &65.8810258056&57.3174845338&56.3980562769&47.4224892693&42.1860841170&45.2876260714 \\
  51 &65.8810258150&57.3174845669&56.3980563285&47.4224893538&42.1860841175&45.2876261744 \\
  61 &65.8810258146&57.3174845661&56.3980563217&47.4224893521&42.1860841144&45.2876261671 \\
  71 &65.8810258149&57.3174845668&56.3980563189&47.4224893537&42.1860841158&45.2876261626 \\
  81 &65.8810258149&57.3174845670&56.3980563187&47.4224893541&42.1860841159&45.2876261622 \\
\hline 
\end{tabular}
\end{center}
\vspace{0.4cm}
{\small {\bf Table 1d:} \quad
Same as Table 1d but for $5^{\rm th}$ order Lagrange FEM.
}
\end{table}
\vskip 0.4cm
\begin{table}[H]
\begin{center}
\begin{tabular}{|l|l|l|l|l|l|l|}
\multicolumn{7}{l}{\bf 6. Order} \\
\hline\hline 
           &             &             &             &             &             & \\
 $n_{nod}$ & $E_{1s1/2}$ & $E_{1p3/2}$ & $E_{1p1/2}$ & $E_{1d5/2}$ & $E_{2s1/2}$ & 
 $E_{1d3/2}$  \\
\hline\hline
   7  & 65.8637153942&57.2828409504&56.3605696991&47.3461657413&41.7817417812&45.2297330851\\
  13  & 65.8808951903&57.3169102490&56.3973587036&47.4203213172&42.1844043482&45.2854747005\\
  19  & 65.8810243462&57.3174818080&56.3980718346&47.4224854807&42.1860829529&45.2876435014\\
  25  & 65.8810248977&57.3174812571&56.3980529004&47.4224808348&42.1860825374&45.2876204571\\
  31  & 65.8810257970&57.3174845238&56.3980563858&47.4224892554&42.1860840306&45.2876261153\\
  37  & 65.8810257899&57.3174845033&56.3980564208&47.4224892313&42.1860839889&45.2876263281\\
  43  & 65.8810258112&57.3174845559&56.3980563261&47.4224893294&42.1860841068&45.2876261751\\
  49  & 65.8810258148&57.3174845666&56.3980563187&47.4224893531&42.1860841163&45.2876261622\\
  55  & 65.8810258149&57.3174845670&56.3980563189&47.4224893542&42.1860841157&45.2876261625\\
  61  & 65.8810258149&57.3174845670&56.3980563187&47.4224893541&42.1860841158&45.2876261623\\
  67  & 65.8810258149&57.3174845670&56.3980563187&47.4224893541&42.1860841158&45.2876261623\\
  73  & 65.8810258149&57.3174845670&56.3980563187&47.4224893541&42.1860841159&45.2876261622\\
\hline 
\end{tabular}
\end{center}
\vspace{0.4cm}
{\small {\bf Table 1e:} \quad
Same as Table 1d but for $6^{\rm th}$ order Lagrange FEM.
}
\end{table}
\vskip 0.4cm
\begin{table}[H]
\begin{center}
\begin{tabular}{|l|l|l|l|l|l|l|}
\multicolumn{7}{l}{\bf 7. Order} \\
\hline\hline 
           &             &             &             &             &             & \\
 $n_{nod}$ & $E_{1s1/2}$ & $E_{1p3/2}$ & $E_{1p1/2}$ & $E_{1d5/2}$ & $E_{2s1/2}$ & 
 $E_{1d3/2}$  \\
\hline\hline
   8 &65.8665180817&57.2917538980&56.3821989755&47.4049510540&42.0593121642&45.2786107904 \\
  15 &65.8808831962&57.3171483037&56.3981295386&47.4219539180&42.1850478466&45.2877364182 \\
  22 &65.8810247273&57.3174817834&56.3980551153&47.4224839062&42.1860817524&45.2876261615 \\
  29 &65.8810255033&57.3174838576&56.3980570146&47.4224881798&42.1860821285&45.2876272423 \\
  36 &65.8810258102&57.3174845542&56.3980563460&47.4224893286&42.1860840744&45.2876262005 \\
  43 &65.8810258149&57.3174845669&56.3980563190&47.4224893538&42.1860841147&45.2876261627 \\
  50 &65.8810258149&57.3174845670&56.3980563186&47.4224893540&42.1860841156&45.2876261621 \\
  57 &65.8810258149&57.3174845670&56.3980563187&47.4224893541&42.1860841158&45.2876261622 \\
  64 &65.8810258149&57.3174845670&56.3980563187&47.4224893541&42.1860841158&45.2876261622 \\
  71 &65.8810258149&57.3174845670&56.3980563187&47.4224893541&42.1860841159&45.2876261622 \\
\hline
\end{tabular}
\end{center}
\vspace{0.4cm}
{\small {\bf Table 1f:} \quad
Same as Table 1e but for $7^{\rm th}$ order Lagrange FEM.
}
\end{table}
\vskip 0.4cm
%
%
%
%

%
%
%
Table 1d displays eigenvalues which have been 
calculated with finite elements of $5^{\rm th}$ order.
At 76 mesh points 12 digits have stabilized 
for all 6 eigenvalues. At 41 mesh points the
precision is already as good as the precision 
in Table 1a at 600 mesh points. A comparison
of the eigenvalues in Table 1d with results of 
a $6^{\rm th}$ order FEM calculation in Table 1e shows
that a further increase of the order leads to 
a rather weak reduction of the number of
required mesh points. At least 73 mesh points 
are necessary in $6^{\rm th}$ order for a precision of
12 digits. As shown in Table 1f, the reduction
of the number of mesh points is even weaker
when the order is increased from $6^{\rm th}$ order to
$7^{\rm th}$ order.
In the subsequent Tables 2a to 2f, results of
corresponding calculations with B-spline finite
elements are shown. In Table 2a, neutron single
particle eigenvalues are listed which have been
calculated with the new B-spline FEM code.
A comparison of the numbers with those listed in Table 1a
shows that they are identical for equal mesh point numbers.
For increasing order of the B-splines, the number of
required mesh points to obtain a certain precision reduces
very similarly to the trend observed in the Tables 1a
to 1f. A comparison of the Tables 2a to 2f with the 
corresponding Tables 1a to 1f shows that roughly half
the number of mesh points is required in a B-spline FEM
in order to achieve the precision of a corresponding
calculation with Lagrangian finite elements. 
In Table 1b, full precision is achieved at 60 mesh points
while 121 mesh points were necessary in Table 1b.
In a calculation with $4^{\rm th}$ order B-spline elements,
45 mesh points are required as shown in Table 2c whereas
145 mesh points are necessary with Lagrange elements (Table 1c)
to obtain a precision of 12 digits.
In the $5^{\rm th}$ order B-spline FEM, 34 mesh points have been used
(Table 2d) while a corresponding $5^{\rm th}$ order Langrange FEM
required 76 mesh points (Table 1d). The $6^{\rm th}$ order B-spline
FEM (see results in Table 2e) leads still to a considerable
relative reduction of the number of mesh points from 34
to 30 at the same level of precision while in the $7^{\rm th}$ order
method still 29 mesh points were required (Table 2f).
The results shown in the Tables 1a to 1f and in the Tables 2a to
2f lead to the conclusion that the B-spline FEM has its 
optimum at $6^{\rm th}$ order whereas the optimal order of the
Lagrange FEM is at $5^{\rm th}$ order. However, the optimal order
may depend on the required precision.
%
%
%
%
%
%
%
%
{ \small
\begin{table}[H]
\begin{center}
\begin{tabular}{|l|l|l|l|l|l|l|}
\multicolumn{7}{l}{\bf 1. Order} \\
\hline\hline 
           &             &             &             &             &             & \\
 $n_{nod}$ & $E_{1s1/2}$ & $E_{1p3/2}$ & $E_{1p1/2}$ & $E_{1d5/2}$ & $E_{2s1/2}$ & 
 $E_{1d3/2}$  \\
\hline\hline 
  10 & 65.88844974 &57.33594611 &56.42536540 &47.45864481 &42.33620819 &45.34520938  \\ 
  20 & 65.88133752 &57.31827220 &56.39907816 &47.42408909 &42.19281200 &45.28981141  \\
  30 & 65.88108502 &57.31763456 &56.39825020 &47.42279518 &42.18737854 &45.28804244  \\
  40 & 65.88104432 &57.31753150 &56.39811692 &47.42258515 &42.18648990 &45.28775645  \\
  50 & 65.88103335 &57.31750369 &56.39808101 &47.42252842 &42.18624969 &45.28767927  \\
  60 & 65.88102944 &57.31749377 &56.39806819 &47.42250815 &42.18616380 &45.28765171  \\
  70 & 65.88102777 &57.31748952 &56.39806272 &47.42249948 &42.18612708 &45.28763993  \\
  80 & 65.88102696 &57.31748747 &56.39806006 &47.42249529 &42.18610928 &45.28763422  \\
  90 & 65.88102653 &57.31748638 &56.39805866 &47.42249306 &42.18609982 &45.28763119  \\
  100& 65.88102628 &57.31748575 &56.39805785 &47.42249178 &42.18609441 &45.28762946  \\
  150 &65.88102591 &57.31748480 &56.39805662 &47.42248983 &42.18608615 &45.28762681 \\
  200 &65.88102584 &57.31748464 &56.39805641 &47.42248951 &42.18608476 &45.28762637  \\
  300 &65.88102582 &57.31748458 &56.39805634 &47.42248938 &42.18608424 &45.28762620  \\
\hline
\end{tabular}
\end{center}
\vspace{0.4cm}
{\small {\bf Table 2a:} \quad
Same as Table 1a but for B-spline FEM.
}
\end{table}
\begin{table}[H]
\begin{center}
\begin{tabular}{|l|l|l|l|l|l|l|}
\multicolumn{7}{l}{\bf 3. Order} \\
\hline\hline 
           &             &             &             &             &             & \\
 $n_{nod}$ & $E_{1s1/2}$ & $E_{1p3/2}$ & $E_{1p1/2}$ & $E_{1d5/2}$ & $E_{2s1/2}$ & 
 $E_{1d3/2}$  \\
\hline\hline 
  10 &65.88136334 &57.31813399 &56.40016227 &47.42348454 &42.18920652 &45.29093938  \\
  20 &65.88102627 &57.31748559 &56.39805814 &47.42249117 &42.18608805 &45.28762926  \\
  30 &65.88102583 &57.31748459 &56.39805636 &47.42248941 &42.18608422 &45.28762623  \\
  40 &65.88102582 &57.31748457 &56.39805632 &47.42248936 &42.18608413 &45.28762617  \\
  50 &65.88102582 &57.31748457 &56.39805632 &47.42248935 &42.18608412 &45.28762616  \\
  60 &65.88102581 &57.31748457 &56.39805632 &47.42248935 &42.18608412 &45.28762616  \\
  70 &65.88102581 &57.31748457 &56.39805632 &47.42248935 &42.18608412 &45.28762616  \\
\hline 
\end{tabular}
\end{center}
\vspace{0.4cm}
{\small {\bf Table 2b:} \quad
Same as Table 1b but for B-spline FEM.
}
\end{table}
\begin{table}[H]
\begin{center}
\begin{tabular}{|l|l|l|l|l|l|l|}
\multicolumn{7}{l}{\bf 4. Order} \\
\hline\hline 
           &             &             &             &             &             & \\
 $n_{nod}$ & $E_{1s1/2}$ & $E_{1p3/2}$ & $E_{1p1/2}$ & $E_{1d5/2}$ & $E_{2s1/2}$ & 
 $E_{1d3/2}$  \\
\hline\hline

  6 & 65.88135693  &57.31850571  &56.40626398  &47.42273069  &42.18460453  &45.29173241 \\
  7 & 65.87919142  &57.31238878  &56.39343856  &47.41226119  &42.18527059  &45.28088247 \\
  8 & 65.88036297  &57.31616474  &56.40144465  &47.42037951  &42.18049023  &45.29330556 \\
  9 & 65.88118037  &57.31768604  &56.39894718  &47.42267251  &42.18838460  &45.28845841 \\
 10 & 65.88085949  &57.31705080  &56.39814878  &47.42162032  &42.18564082  &45.28786499 \\
 11 & 65.88104483  &57.31749502  &56.39864079  &47.42246069  &42.18618288  &45.28849993 \\
 20 & 65.88102585  &57.31748464  &56.39805656  &47.42248949  &42.18608449  &45.28762653 \\
 25 & 65.88102582  &57.31748457  &56.39805633  &47.42248937  &42.18608414  &45.28762618 \\
 30 & 65.88102582  &57.31748457  &56.39805632  &47.42248936  &42.18608412  &45.28762616 \\
 35 & 65.8810258150&57.3174845672&56.3980563189&47.4224893544&42.1860841164&45.2876261626 \\
 40 & 65.8810258149&57.3174845670&56.3980563187&47.4224893542&42.1860841160&45.2876261623 \\
 45 & 65.8810258149&57.3174845670&56.3980563187&47.4224893541&42.1860841159&45.2876261622 \\
 50 & 65.8810258149&57.3174845670&56.3980563187&47.4224893541&42.1860841159&45.2876261622 \\
\hline 
\end{tabular}
\end{center}
\vspace{0.4cm}
{\small {\bf Table 2c:} \quad
Same as Table 1c but for B-spline FEM.
}
\end{table}
\begin{table}[H]
\begin{center}
\begin{tabular}{|l|l|l|l|l|l|l|}
\multicolumn{7}{l}{\bf 5. Order} \\
\hline\hline 
           &             &             &             &             &             & \\
 $n_{nod}$ & $E_{1s1/2}$ & $E_{1p3/2}$ & $E_{1p1/2}$ & $E_{1d5/2}$ & $E_{2s1/2}$ & 
 $E_{1d3/2}$  \\
\hline\hline
   7 &65.88022738  &57.31648520  &56.40099275  &47.42158367  &42.17900675  &45.29112016  \\
  10 &65.88103149  &57.31749002  &56.39852712  &47.42249777  &42.18607033  &45.28816996  \\
  20 &65.88102581  &57.31748455  &56.39805651  &47.42248931  &42.18608408  &45.28762647  \\
  30 &65.8810258149&57.3174845671&56.3980563188&47.4224893542&42.1860841160&45.2876261624 \\
  31 &65.8810258149&57.3174845670&56.3980563187&47.4224893542&42.1860841160&45.2876261623 \\
  32 &65.8810258149&57.3174845670&56.3980563187&47.4224893542&42.1860841159&45.2876261623 \\
  33 &65.8810258149&57.3174845670&56.3980563187&47.4224893542&42.1860841159&45.2876261622 \\
  34 &65.8810258149&57.3174845670&56.3980563187&47.4224893541&42.1860841159&45.2876261622 \\
\hline 
\end{tabular}
\end{center}
\vspace{0.4cm}
{\small {\bf Table 2d:} \quad
Same as Table 1d but for B-spline FEM.
}
\end{table}
\begin{table}[H]
\begin{center}
\begin{tabular}{|l|l|l|l|l|l|l|}
\multicolumn{7}{l}{\bf 6. Order} \\
\hline\hline 
           &             &             &             &             &             & \\
 $n_{nod}$ & $E_{1s1/2}$ & $E_{1p3/2}$ & $E_{1p1/2}$ & $E_{1d5/2}$ & $E_{2s1/2}$ & 
 $E_{1d3/2}$  \\
\hline\hline

  10 &65.88092534  &57.31722378  &56.39803360  &47.42198939  &42.18580853  &45.28768763  \\
  15 &65.88102458  &57.31748100  &56.39805735  &47.42248186  &42.18608340  &45.28762800  \\
  20 &65.8810258052&57.3174845391&56.3980563477&47.4224892980&42.1860841341&45.2876262023 \\
  25 &65.8810258149&57.3174845669&56.3980563193&47.4224893539&42.1860841164&45.2876261632 \\ 
  26 &65.8810258148&57.3174845668&56.3980563194&47.4224893537&42.1860841155&45.2876261633 \\
  27 &65.8810258149&57.3174845671&56.3980563188&47.4224893542&42.1860841159&45.2876261625 \\
  28 &65.8810258149&57.3174845670&56.3980563188&47.4224893541&42.1860841159&45.2876261623 \\
  29 &65.8810258149&57.3174845670&56.3980563187&47.4224893541&42.1860841159&45.2876261623 \\
  30 &65.8810258149&57.3174845670&56.3980563187&47.4224893541&42.1860841159&45.2876261622 \\
\hline
\end{tabular}
\end{center}
\vspace{0.4cm}
{\small {\bf Table 2e:} \quad
Same as Table 1e but for B-spline FEM.
}
\end{table}
\begin{table}[H]
\begin{center}
\begin{tabular}{|l|l|l|l|l|l|l|}
\multicolumn{7}{l}{\bf 7. Order} \\
\hline\hline 
           &             &             &             &             &             & \\
 $n_{nod}$ & $E_{1s1/2}$ & $E_{1p3/2}$ & $E_{1p1/2}$ & $E_{1d5/2}$ & $E_{2s1/2}$ & 
 $E_{1d3/2}$  \\
\hline\hline

  10 &65.88099742  &57.31745060  &56.39818784  &47.42246110  &42.18582288  &45.28773718  \\
  15 &65.88102496  &57.31748290  &56.39806113  &47.42248693  &42.18607611  &45.28763254  \\
  20 &65.88102580  &57.31748453  &56.39805640  &47.42248929  &42.18608399  &45.28762629  \\
  25 &65.8810258147&57.3174845665&56.3980563197&47.4224893532&42.1860841144&45.2876261638 \\
  26 &65.8810258149&57.3174845670&56.3980563188&47.4224893541&42.1860841159&45.2876261623 \\
  27 &65.8810258149&57.3174845669&56.3980563188&47.4224893539&42.1860841157&45.2876261624 \\
  28 &65.8810258149&57.3174845670&56.3980563187&47.4224893541&42.1860841158&45.2876261622 \\
  29 &65.8810258149&57.3174845670&56.3980563187&47.4224893541&42.1860841159&45.2876261622 \\
\hline
\end{tabular}
\end{center}
\vspace{0.4cm}
{\small {\bf Table 2f:} \quad
Same as Table 1f but for B-spline FEM.
}
\end{table}
To complete this study, I repeated the calculations 
for a large number of mesh points with
both methods from $1^{\rm st}$ order to $8^{\rm th}$ order finite elements.
In the figures Fig. 8a to Fig. 8d, the logarithmic errors with respect to the
highest precision are shown. Fig. 8a displays the averaged error 
taken over all 6
single particle energies which have been calculated 
with the B-spline FEM. In Fig. 8b,
these data have been smoothed by taking in addition the average over two 
neighboring mesh point numbers. 
In the region of precision ranging from $10^{-1}$ to
$10^{-10.5}$, both figures display an enormous 
reduction of the errors for increasing
finite element order up to $5^{\rm th}$ order.
Finite elements of higher order do not essentially improve the precision in the
considered range but entail a higher numerical cost. 
They may improve the precision
beyond the error range of $\bigl[ 10^{-1},\, 10^{-10}\bigr]$.
However, precisions in the range of $\bigl[ 10^{-1},\, 10^{-10}\bigr]$
are sufficient in most applications.
In Fig. 8b there is an indication for $6^{\rm th}$ order to become 
optimal order at precisions better than $10^{-10}$. 
This is in agreement with the 
conclusion that has been drawn form 
the data in Table 2a to 2f.

In Fig. 8c and Fig. 8d. results that 
correspond to Fig. 8a and Fig. 8b but calculated
with the Lagrange FEM are displayed. 
A similar but weaker reduction of the errors
is observed for increasing finite element order. 
A comparison with the results
depicted in Fig. 8a and Fig. 8b shows that 
the B-spline method requires in general
a much smaller number of mesh points than the Lagrange FEM 
in order to provide the
desired level of precision. For comparison, in Fig. 8d, 
I have inserted those graphs 
of Fig. 8b which resulted for $5^{\rm th}$ order to $8^{\rm th}$ order B-spline calculations.
%
%
%
%
%
\begin{figure}[H]
\centerline{
{ \epsfysize=6cm \epsfxsize=7cm                  
\epsffile{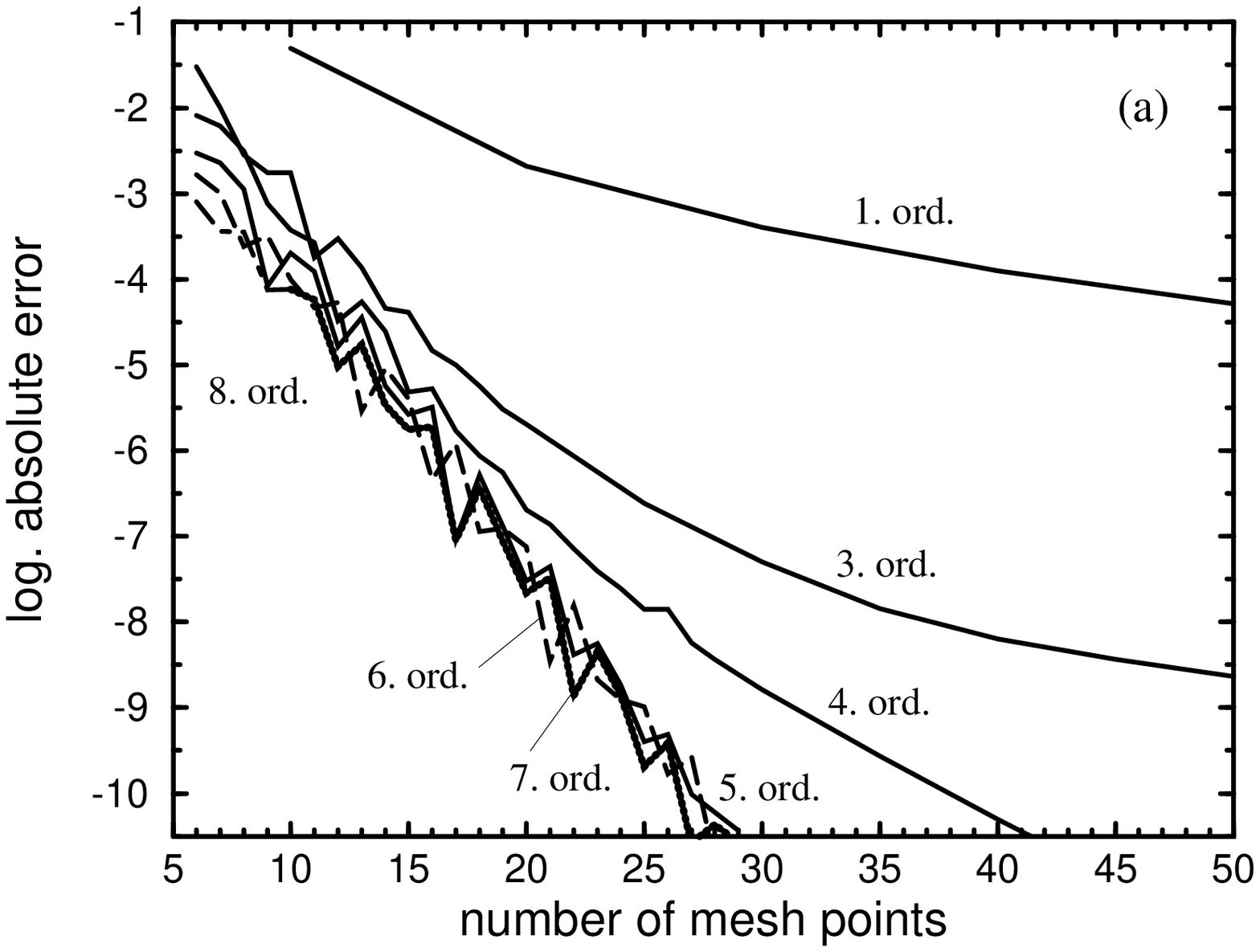} }          \hspace{1cm}
{ \epsfysize=6cm \epsfxsize=7cm                  
\epsffile{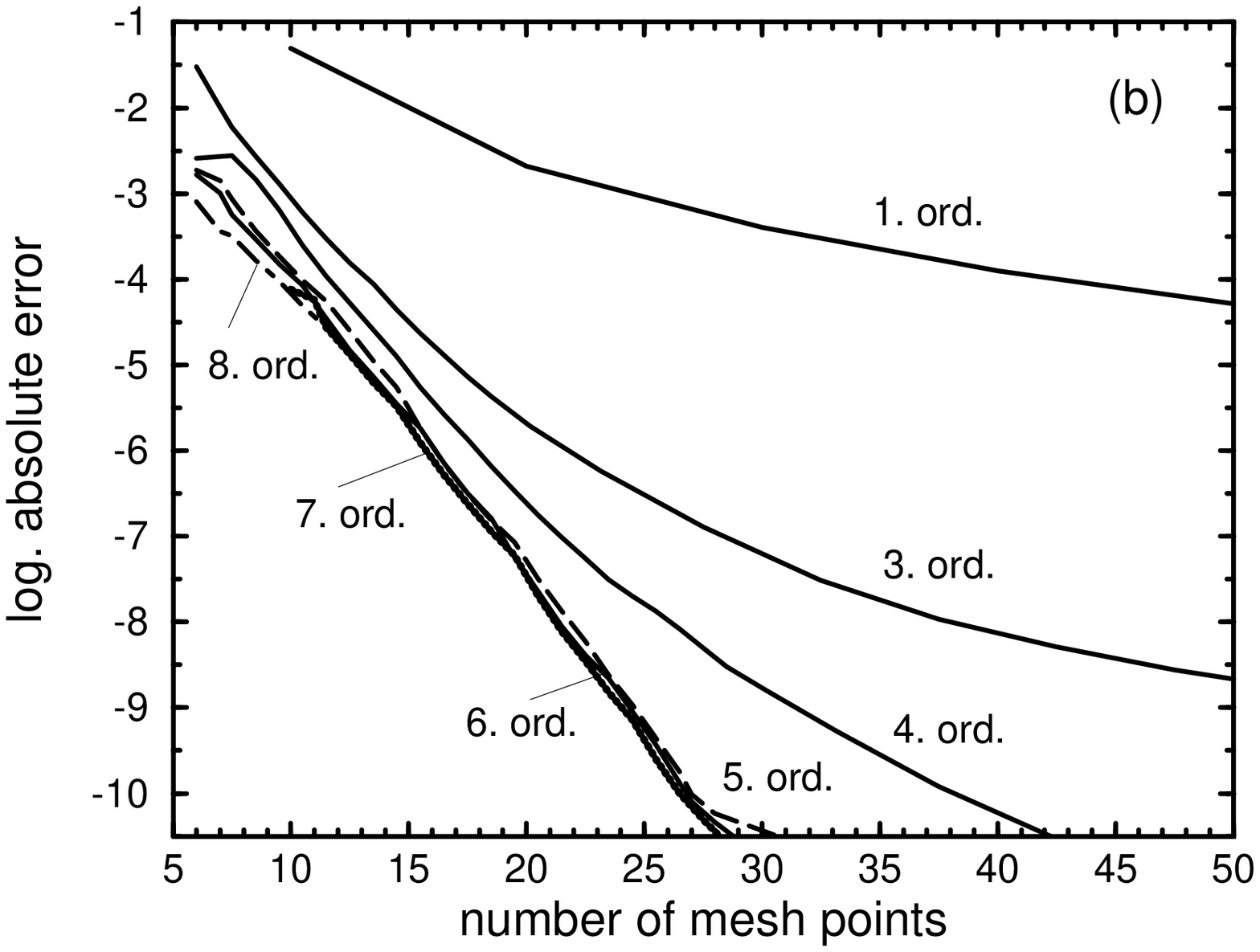} }
}
\end{figure}
\begin{figure}[H]
\centerline{
{ \epsfysize=6cm \epsfxsize=7cm                  
\epsffile{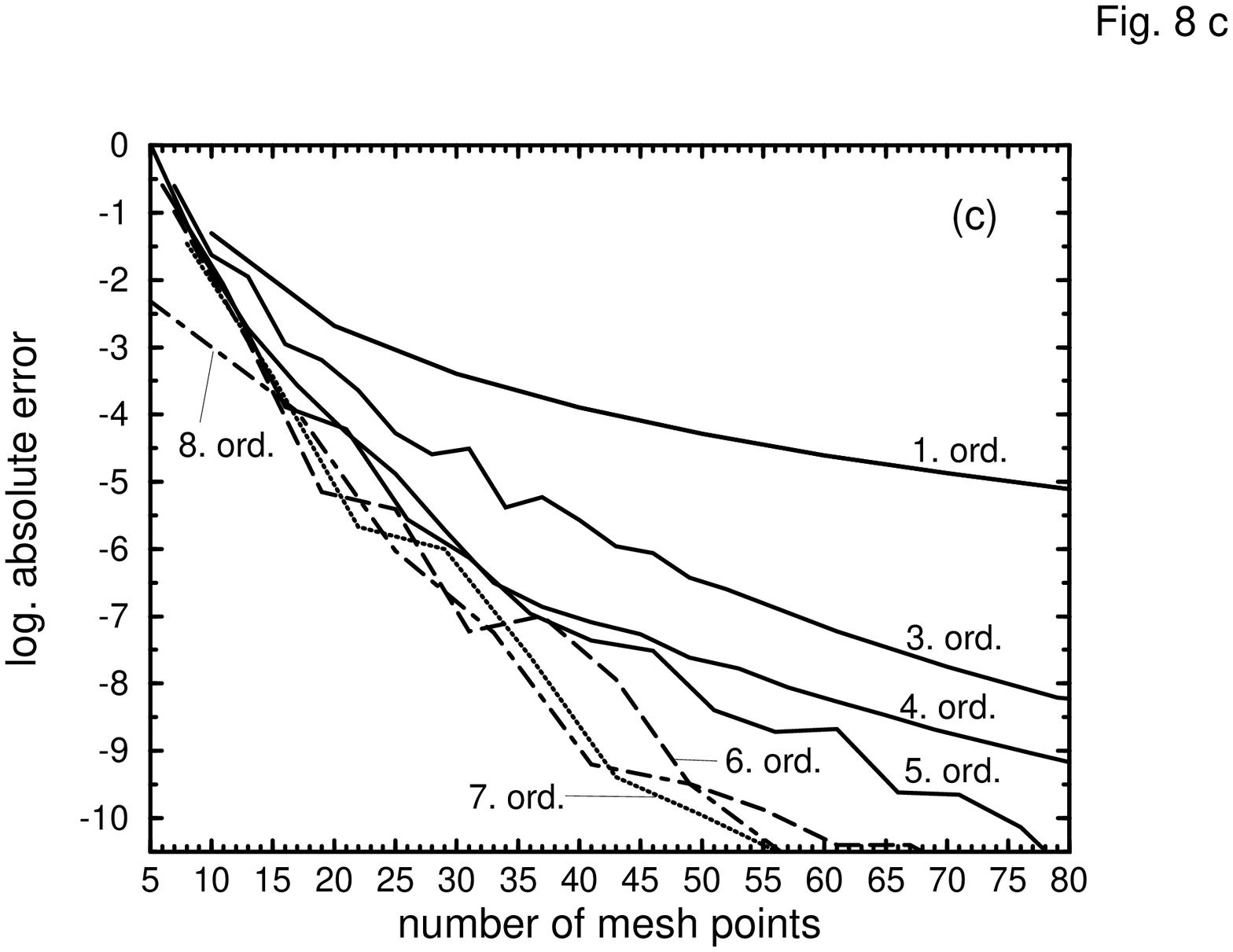} }          \hspace{1cm}
{ \epsfysize=6cm \epsfxsize=7cm                  
\epsffile{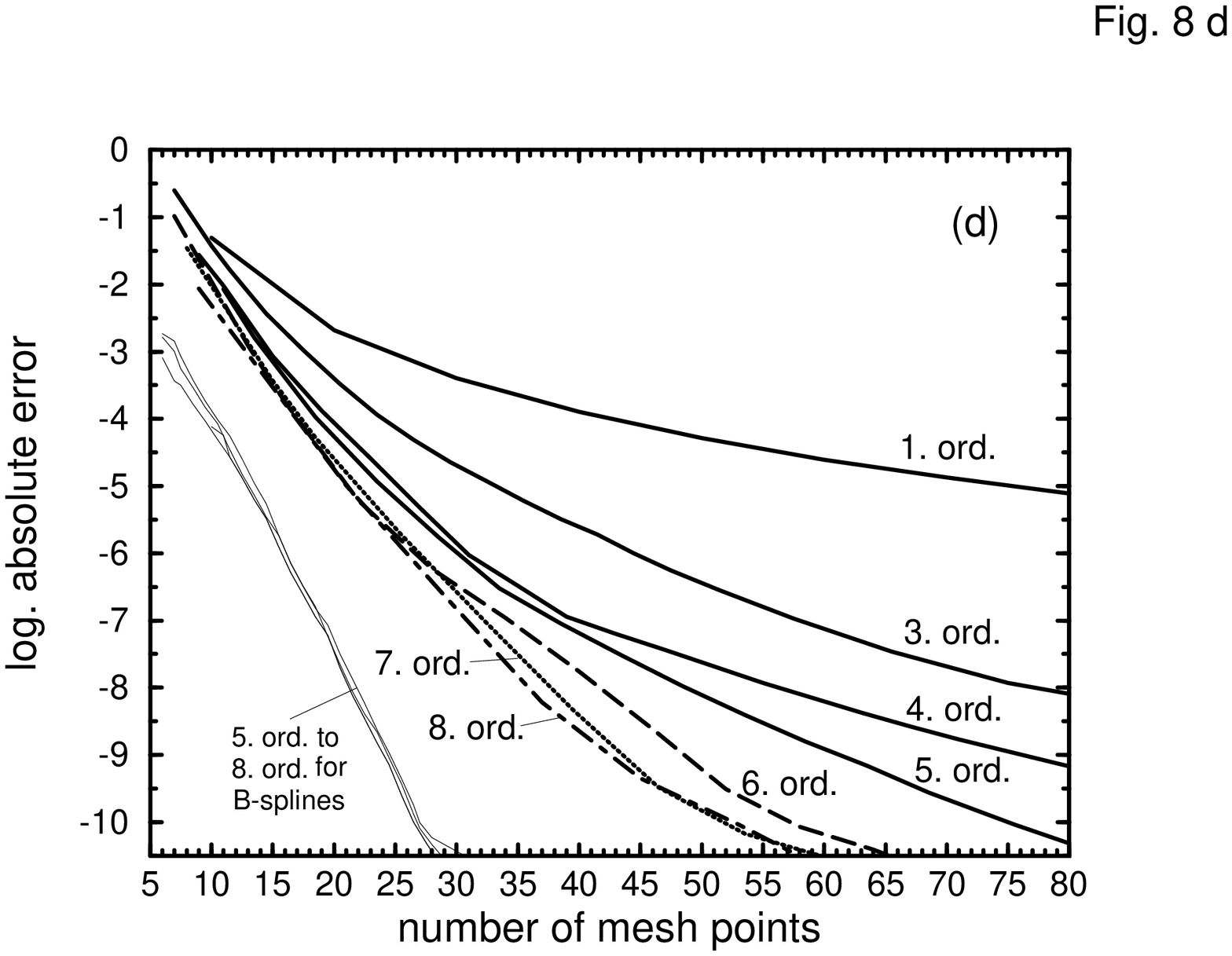} }
}
\vskip 0.5cm    
{\small {\bf Fig. 8:} \quad
Logarithmic plots of the error which occurs
in the B-spline FEM (figures (a),(b))
and in the Lagrange FEM (figures (c),(d)).
The used finite element orders are indicated.
Figs. (a) and (c) display
average values taken over the six lowest positive neutron 
single particle eigenvalues. 
Figs. (b) and (d) show the smoothed curves.
}
\end{figure}
\vskip 0.5cm
%
%
\section{Performance discussion}
%
%
With respect to applications of the above presented B-spline FEM in
large scale computations with FEM discretizations in more than one
dimension, attention should be payed to the performance of both
methods at the present stage. Therefore, I investigate and compare
the run time for both methods, the B-spline FEM and the Lagrange
FEM.
The following discussion is based on data which correspond to the
performance of the codes on a DEC Alpha 300\,MHz. A gnu compiler
has been used under UNIX to translate the codes.

The CPU-time depends essentially on the FEM order
and the number of used mesh points. The times which are displayed in
the Figs. 9a and 9b correspond to a single step in which the Dirac
equation \ref{Equ.2.1} is solved. This procedure encloses 
essentially the construction and the solution of the generalized 
eigenvalue problem for one $\kappa$-value.
It has to be repeated for each $\kappa$ in the solver for
neutron and proton states and this again over the whole 
self-consistent iteration.
The CPU-time which is required for the solution of the meson field 
equations lies below one percent
of that for the nucleon states and is therefore neglected. 
Fig. 9a displays the CPU-time for $5^{\rm th}$ order FEM as a function
of the number of mesh points. 
Results which have been obtained for other orders
are almost identical with those shown in the figure. 
The solid curve displays CPU-times resulting
from Lagrange type finite element discretizations while the
dashed curve has been obtained with the B-spline FEM. 
An explanation for the higher numerical cost in applications of the
B-spline method is given by Figs. 4 showing that more matrix
elements have to be calculated in the case of B-spline FEM.

However, as demonstrated in the Figs. 8, the number of
required mesh points for equal numerical precision is half of that in
the Lagrange FEM. At equal numerical precision, the solid curve has to
be compared with the dot-dashed curve of a B-spline calculation. 
This clearly demonstrates an enormous reduction of the numerical cost
for the B-spline FEM.

In Fig. 9b, CPU-times are plotted as a function of the FEM order
and compared for both methods.
The precision of the numerical solution which has been obtained
with the B-spline FEM
(dashed line) is much higher that that obtained with the Lagrange
FEM (solid line). At equal numerical precision, one should compare
values of the dashed line with values of the solid line at double
order.  
\begin{figure}[H]
\centerline{
{ \epsfysize=6cm \epsfxsize=7cm                  
\epsffile{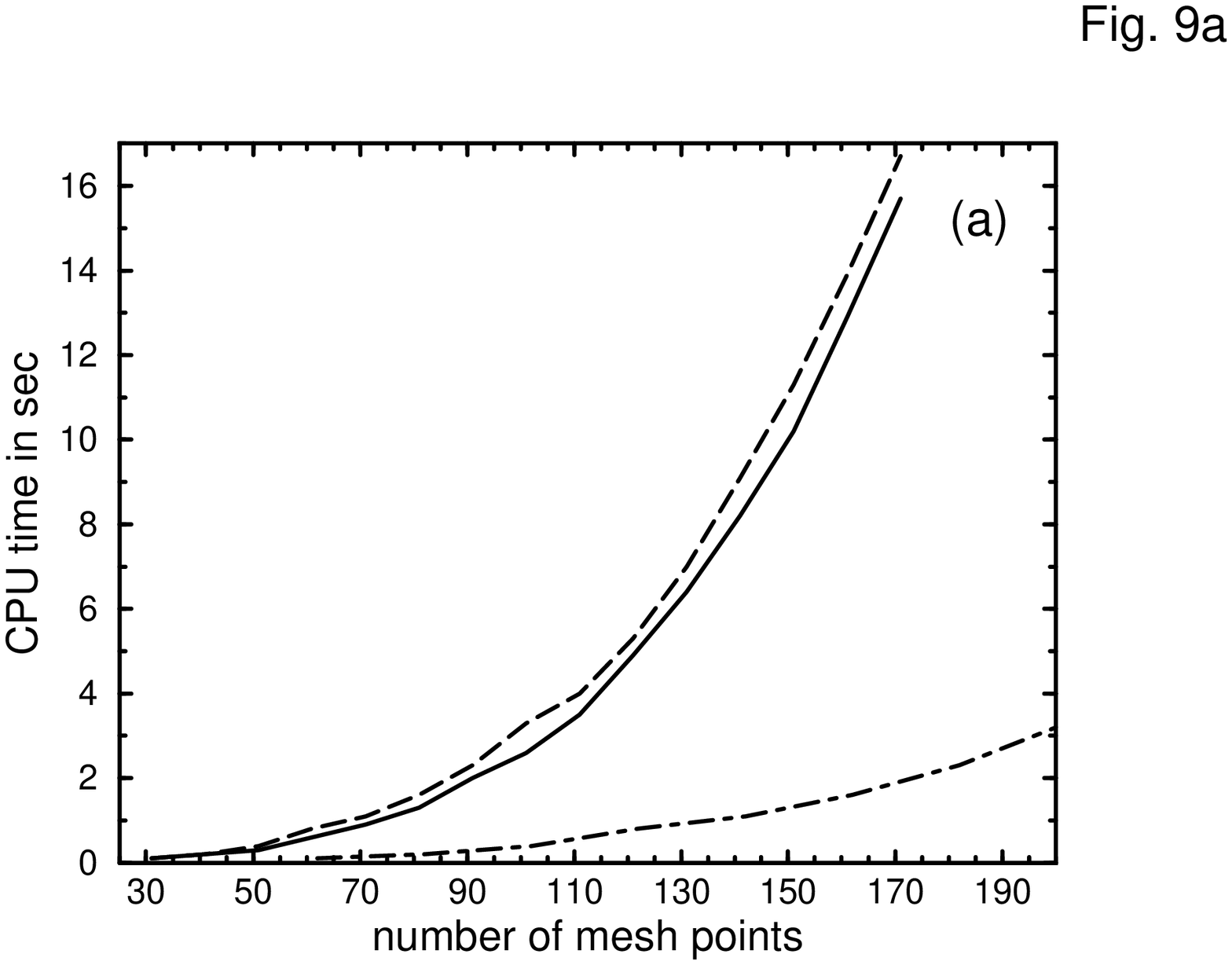} }          \hspace{1cm}
{ \epsfysize=6cm \epsfxsize=7cm                  
\epsffile{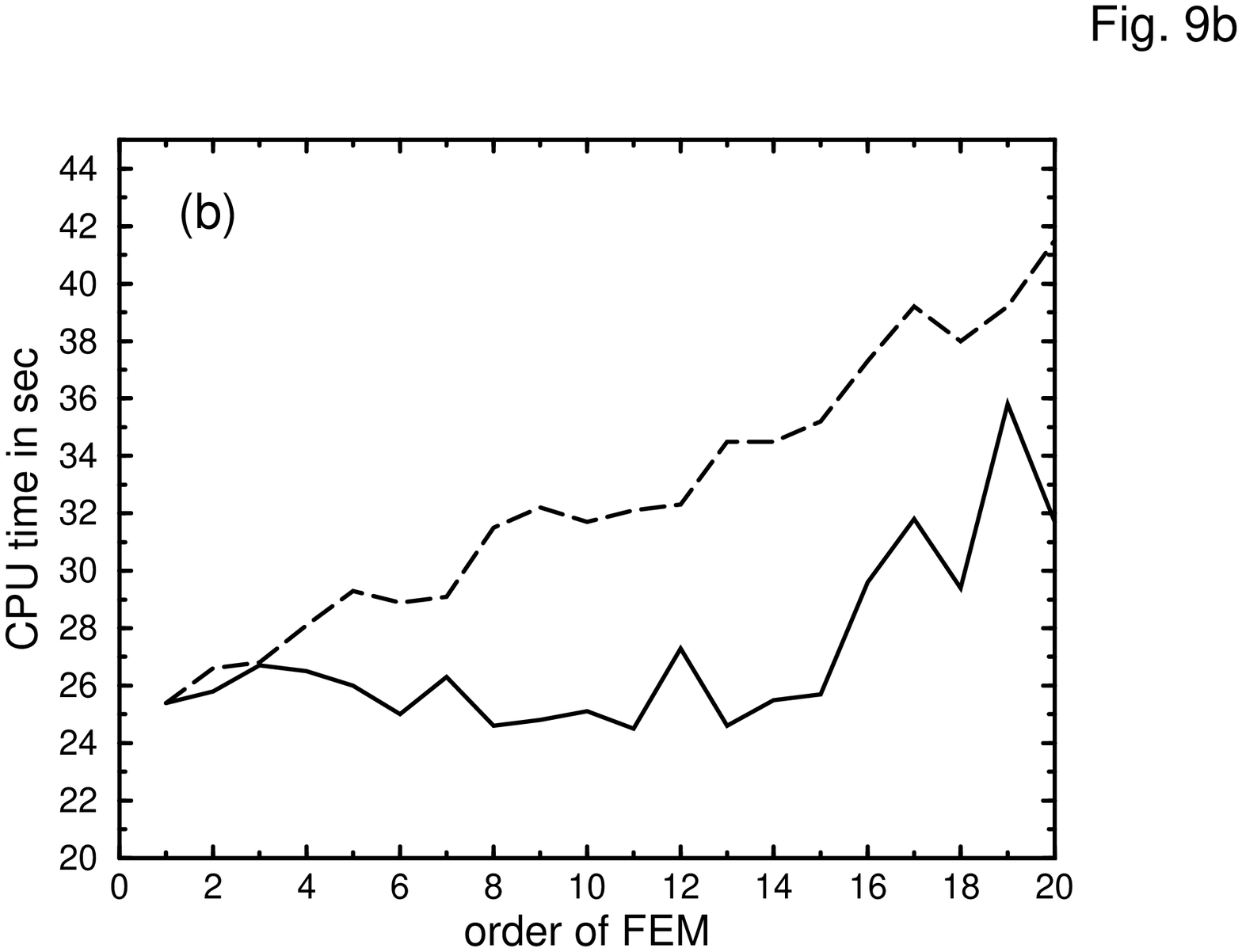} }
}
\vskip 0.5cm    
{\small {\bf Fig. 9:} \quad
CPU time as a function of the number of used mesh points (a) and
as a function of used order (b) for both methods, B-spline FEM
and Lagrange FEM.
}
\end{figure}
%
%
%
%
%
\section{Program structure}
%
The program is coded in C++. The implementation of the relativistic
mean field model in the Hartree approximation for spherical doubly-closed
shell nuclei has been described in Ref.~\cite{PVRR.97}. 
In this section we only describe the changes that have 
been made in order to modify the program to B-spline techniques.

The main part of the program consists of seven classes:
{\bf MathPar}: numerical parameters used in the code,
{\bf PhysPar}: physical parameters (masses,
coupling constants, etc.),
{\bf FinEl}: finite elements, 
{\bf Mesh}: mesh in coordinate space,
{\bf Nucleon}: neutrons and protons in
the nuclear system, {\bf Meson}: mesons and photon with corresponding
mean fields and the Coulomb field, and the class {\bf LinBCGOp}.
A detailed description of these classes can be found in Ref.~\cite{PVRR.97}.

Two new classes {\bf FinElBsp} 
and {\bf BSpline} 
have been added to the code.
The implementation can be found in the source files 
finelbsp.cc, bspline.cc and the corresponding
header files finelbsp.h, bspline.h. 
The class {\bf FinElBsp} 
contains the following new member functions: \hfill\break
{\it
FinElBsp();
\hfill\break
\~FinElBsp();
\hfill\break
void FinElBsp::alloc( int ord );                        \hfill\break
void FinElBsp::free();                                  \hfill\break
void FinElBsp::make( int ord );                         \hfill\break
double FinElBsp::n( int iloc, int ife, int iga, int l, bool zero ) const;
\hfill\break
double FinElBsp::dn( int iloc, int ife, int iga, int l, bool zero ) const;
\hfill\break
double FinElBsp::func( double const* u, int ife, int iga ) const;
\hfill\break
double FinElBsp::func( double const* u,int ife,int iga,int l,bool zero ) const;
\hfill\break
void FinElBsp::eval();
\hfill\break
inline int FinElBsp::order() const;
\hfill\break
inline int FinElBsp::nloc() const;
\hfill\break
inline double FinElBsp::n( int iloc, int iga ) const;
\hfill\break
inline double FinElBsp::dn( int iloc, int iga ) const;
\hfill\break
inline double FinElBsp::operator()(double const* u,int ife,int iga) const;
\hfill\break
inline double FinElBsp::operator()(double const* u,int ife,int iga,int l, 
                                   bool zero ) const;
}
The two methods {\it FinElBsp()} and {\it \~FinElBsp()} describe the
constructor and destructor of objects (B-spline finite elements) of the class.
The method {\it make( int ord ) } provids the B-spline reference element
with data (amplitudes) on the shape functions and their derivatives.
Access to these data is given through the methodes
{\it n( int iloc, int iga )} and {\it dn( int iloc, int iga ) }.
In a first step, {\it make( int ord ) } allocates memory for the
shape functions using {\it alloc( int ord )}.
In a second step, the method {\it eval()} is called generating
the amplitudes of the shape functions through an operator of class
{\bf BSpline }.
The overload member functions
{\it n( int iloc, int ife, int iga, int l, bool zero ) const }
and
{\it dn( int iloc, int ife, int iga, int l, bool zero ) const}
are used in the calculation of the stiffness matrices. 
They take into account boundary conditions. The method
{\it func( double const* u, int ife, int iga ) const}
provides the interpolated amplitude of solutions on the Gauss submesh
in any finite element of global index {\it ife}.  
The overloaded version
{\it func( double const* u, int ife, int iga ) const}
takes boundary conditions into account.

In the class {\bf BSpline}, the following methods are implemented
{ \it
BSpline( int ord ) \hfill\break
\~BSpline()        \hfill\break
operator()( double const* p, double\& f, double\& df, double x ) \hfill\break
inline int BSpline::order() const; \hfill\break
}
{\it BSpline( int ord )} and 
{\it \~BSpline()       } describe the constructor and the
destructor of the class.
An operator is used to carry out the B-spline algorithm (\ref{Equ.3.10})
whenever access to B-spline amplitudes is requested through a call
of an object of the class with corresponding arguments.

The organization in the construction of stiffness matrices in other
parts of the code has been changed accordingly. However, the essential
structure has been maintained so that quick changes for applications
of Langrange type finite elements (defined in class {\bf FinEl}) are
possible. An essential difference roots in the relation between 
number of nodes on the global mesh and the number of finite elements
which is given by $n^{\rm nod}=n^{\rm fe}+n^{\rm ord}$.
In the version using Langrange type elements this relation is
$n^{\rm nod}=n^{\rm fe}\cdot n^{\rm ord} + 1$.
 
For the diagonalization of the generalized eigenvalue problems the
bisection method has been replaced by a combined
Cholesky decomposition and householder method which
is slower but allows for a higher precision. 
A heapsort algorithm orders eigenvalues and eigenvectors.
The routines and the eigensolver are implemented in the
source file eigen.cc and the header file eigen.h.

\vskip 2.0 cm
\appendix
%
\section*{ Appendix}
\vskip0.5cm
%
For the FEM discretization of the Klein-Gordon equations  
we use the ansatz
\begin{eqnarray}
\label{App.4.8.a}
\sigma (r)=\sum\limits_p \sigma_p  B_p(r) \, \\
\label{App.4.8.b}
\omega^0 (r)=\sum\limits_p \omega^0_p  B_p(r) \, \\
\label{App.4.8.c}
\rho^0 (r)=\sum\limits_p \rho^0_p  B_p(r) \, \\
\label{App.4.8.d}
A^0 (r)=\sum\limits_p A^0_p  B_p(r).
\end{eqnarray}
where the node variables $\sigma_p,$ $\omega^0_p,$ $\rho^0_p$ and
$A^0_p$ correspond to field amplitudes at the mesh point p.
For the Klein-Gordon equations we use the same type of
shape functions $B_p(r)$, and the same mesh as in the 
FEM discretization of the Dirac equation (\ref{Equ.4.2}). 
Using again the method of weighted residuals with test functions
$w_p(r)=r^2B_p(r)$, the following algebraic equations are obtained 
\begin{eqnarray}
\sum\limits_p\Bigl< w_{p'}(r)\Big\vert
 -\partial_r^2-{2\over r}\partial_r+{{l(l+1)}\over{r^2}}+
m_{\sigma}^2 \Big\vert B_p(r)\Bigr>
\sigma_p = \Bigl< w_{p'}(r)\Big\vert s_{\sigma}(\Phi_1(r),...,\Phi_A(r))\Bigr>
\, \\
\sum\limits_p\Bigl< w_{p'}(r)\Big\vert
-\partial_r^2-{2\over r}\partial_r+{{l(l+1)}\over{r^2}}+
m_{\omega}^2 
\Big\vert B_p(r)\Bigr>
\omega^0_p = 
\Bigl< w_{p'}(r)\Big\vert s_{\omega}(\Phi_1(r),...,\Phi_A(r))\Bigr>
\, \\
\sum\limits_p\Bigl< w_{p'}(r)\Big\vert
 -\partial_r^2-{2\over r}\partial_r+{{l(l+1)}\over{r^2}}+
m_{\rho}^2 
\Big\vert B_p(r)\Bigr>
\rho^0_p = \Bigl< w_{p'}(r)\Big\vert s_{\rho}(\Phi_1(r),...,\Phi_A(r))\Bigr>
\, \\
\sum\limits_p\Bigl< w_{p'}(r)\Big\vert
 -\partial_r^2-{2\over r}\partial_r+{{l(l+1)}\over{r^2}}
\Big\vert B_p(r)\Bigr>
A^0_p = \Bigl< w_{p'}(r)\Big\vert s_{C}(\Phi_1(r),...,\Phi_A(r))\Bigr> .
\end{eqnarray}
The resulting matrix equations read
\begin{eqnarray}
\label{App.4.9.a}
\Bigl[S^{\sigma}_1+l(l+1)\cdot S^{\sigma}_2+
m_{\sigma}^2\cdot  S^{\sigma}_3+S^{\sigma}_4\Bigr] {\vec\sigma}\,
& = & {\vec r^{\rm (s)}} \, \\
\label{App.4.9.b}
\Bigl[S^{\omega}_1+l(l+1)\cdot S^{\omega}_2+
m_{\omega}^2\cdot S^{\omega}_3\Bigr]\cdot{\vec\omega^0} & = &
{\vec r^{\rm (v)}} \, \\
\label{App.4.9.c}
\Bigl[S^{\rho}_1+l(l+1)\cdot S^{\rho}_2+
m_{\rho}^2\cdot S^{\rho}_3\Bigr]\cdot{\vec\rho^0} & = &
{\vec r^{\rm (3)}}  \, \\
\label{App.4.9.d}
\Bigl[S^{A^0}_1+l(l+1)\cdot S^{A^0}_2\Bigr]\cdot{\vec A^0} & = &
{\vec r^{\rm (em)}} .
\end{eqnarray}
\noindent
The node variables $\sigma_p,$ $\omega^0_p,$ $\rho^0_p$ and
$A^0_p$ are grouped 
into the vectors ${\vec\sigma}=(\sigma_1,...,\sigma_n)^T$,
${\vec\omega^{\,0}}=(\omega_1,...,\omega_n)^T$, 
${\vec\rho^{\,0}}=(\rho^0_1,...,\rho^0_n)^T$, ${\vec A^0}=(A^0_1,...,A^0_n)^T$,
and \begin{eqnarray}
\label{App.4.10.a}
S_1^{\sigma} = S_1^{\omega} = S_1^{\rho} = S_1^{A} &=&
\Bigl< w_{p'}(r)\Big\vert \partial_r^2 + 2\,r^{-1}\,\partial_r 
\Big\vert B_p(r) \Bigr>,                                       \, \\
\label{App.4.10.b}
S_2^{\sigma} = S_2^{\omega} = S_2^{\rho} = S_2^{A} &=&
\Bigl< w_{p'}(r)\Big\vert r^{-2}\Big\vert B_p(r)\Bigr>,      \, \\
\label{App.4.10.c}
S_3^{\sigma} = S_3^{\omega} = S_3^{\rho} = S_3^{A} &=&
\Bigl< w_{p'}(r)\Big\vert B_p(r)\Bigr>,           \,  \\
\label{App.4.10.d}
S_4^{\sigma} &=& 
\Bigl< w_{p'}(r)\Big\vert g_2\sigma(r)+g_3\sigma(r)^2\Big\vert B_p(r)\Bigr>.
\end{eqnarray}
The components of the right hand side vectors are defined as
\begin{eqnarray}
r^{\rm (s)}_{p'}&=&-g_{\sigma}\Bigl<w_{p'}(r)\Big\vert\rho_{\rm s}(r)\Bigr>, 
\nonumber\\
r^{\rm (v)}_{p'}&=& g_{\omega}\Bigl<w_{p'}(r)\Big\vert\rho_{\rm v}(r)\Bigr>, 
\nonumber\\
r^{(3)}_{p'}&=& g_{\rho}\Bigl<w_{p'}(r)\Big\vert\rho_3(r)\Bigr>, \nonumber\\
r^{\rm (em)}_{p'}&=& e \Bigl<w_{p'}(r)\Big\vert\rho_{\rm em}(r)\Bigr>.
\label{App4.11}
\end{eqnarray}
The nonlinear equation for the $\sigma$-field is solved in an 
iterative procedure.
The nonlinear terms are included in the global stiffness matrix 
(see $S^{\sigma}_4$ in Eq. (\ref{App.4.9.a})).
In the iterative solution matrix elements that contain nonlinear terms are
calculated using the field $\sigma(r)$ obtained in the previous iteration step.
%
%











\end{document}